\documentstyle[11pt,epsf]{article}
\newdimen\mathindent
\renewcommand{\qquad}{\hspace*{25pt}}
\newcommand{\eref}[1]{(\ref{#1})}

\def\wP{\widehat P}
\def\wPm{$\widehat P$-matrix}
\def\wPms{$\widehat P$-matrices}

\def\prodd#1#2#3{\prod\limits_{#1}^{#2}\lower3pt\hbox{${ }_{#3}$}}

\newcommand{\eqll}[2]{\begin{equation}{#1}\label{#2}\end{equation}}

\newcommand{\eq}[1]{\begin{equation}{#1}\end{equation}}
\def\integer{{\mathchoice
    {\hbox{ $\displaystyle\kern-1mm {\rm Z}\kern-1.1mm {\rm Z}$}}
    {\hbox{ $\textstyle\kern-1mm {\rm Z}\kern-1.1mm {\rm Z}$}}
    {\hbox{$\scriptstyle\kern-1mm {\rm Z}\kern-1.1mm {\rm Z}$}}
    {\hbox{$\scriptscriptstyle\kern-1mm {\rm Z}\kern-1.1mm {\rm Z}$}}}}
\def\natural{{\mathchoice
    {\hbox{ $\displaystyle\kern-1.4mm 1\kern-.7mm {\rm N}$}}
    {\hbox{ $\textstyle\kern-1.4mm 1\kern-.7mm {\rm N}$}}
    {\hbox{$\scriptstyle\kern-1.4mm 1\kern-.7mm {\rm N}$}}
    {\hbox{$\scriptscriptstyle\kern-1.4mm 1\kern-.7mm {\rm N}$}}}}
\def\real{{\mathchoice
    {\hbox{$\displaystyle\kern-.2mm  I\!\!R $}}
    {\hbox{$\textstyle\kern-.2mm  I\!\!R $}}
    {\hbox{$\scriptstyle\kern-.2mm  I\!\!R $}}
    {\hbox{$\scriptscriptstyle\kern-.2mm  I\!\!R\kern-.2mm $}}}}
\def\complex{{\mathchoice
    {\hbox{$\displaystyle\kern-.2mm {\rm C}\kern-1.5mm\raise.2mm
                   \hbox{\vrule height6pt}\kern1.3mm$}}
    {\hbox{$\textstyle\kern-.2mm {\rm C}\kern-1.5mm\raise.3mm
                   \hbox{\vrule height6pt}\kern1.3mm$}}
    {\hbox{$\scriptstyle\kern-.2mm{\rm C}\kern-1.5mm\raise.2mm
                   \hbox{\vrule height3pt}\kern1.3mm$}}
    {\hbox{$\scriptscriptstyle\kern-.2mm{\rm C}\kern-1.5mm\raise.2mm
                  \hbox{\vrule height2pt}\kern1.3mm$}}}}
\def\summ#1#2#3{\sum\limits_{#1}^{#2}\lower3pt\hbox{${ }_{#3}$}}
\def\scalarp#1#2{\langle~{#1}~,~{#2}~\rangle}
\def\prodd#1#2#3{\prod\limits_{#1}^{#2}\lower3pt\hbox{${ }_{#3}$}}

\newtheorem{definition}{Definition}[section]

\newtheorem{proposition}{Proposition}[section]

\newtheorem{remark}{Remark}[section]

\newtheorem{theorem}{Theorem}[section]

\newtheorem{lemma}{Lemma}[section]

\newcommand{\scat}{\hfill $\Box$ \vspace{3ex}}

\begin{document}

\bibliographystyle{unsrt}

\title{
%\raggedright
Determination of the orbit spaces of non-coregular compact linear groups with
one relation among the basic polynomial invariants in the $\hat P$-matrix
approach\footnote{This paper is partially supported by INFN and MURST 40\% and
60\%, and is carried out as part of the European Community Program ``Gauge
Theories, Applied Supersymmetry and Quantum Gravity" under contract
SCI-CT92-D789.}}

\author{G. Sartori and G. Valente\\
\small Dipartimento di Fisica, Universit\`a di Padova  \\
\small and INFN, Sezione di Padova \small I--35131 Padova, Italy \\
\small (e-mail: gfsartori@padova.infn.it,
valente@padova.infn.it)}

\date{}

\maketitle

\begin{abstract}
Invariant functions under the transformations of a compact linear group $G$
acting in $\real^n$ can be expressed in terms of functions defined in the
orbit space of $G$, i.e.\ as functions of a finite set of basic invariant
polynomials $p(x)=(p_1(x ),\dots ,p_q(x))$, $x\in\real^n$, which form an
integrity basis (IB) for $(G, \real^n)$.

We develop a method to determine the isotropy classes of the orbit spaces
of all the real linear groups whose IBs satisfy only one independent
relation. The effectiveness of the method is tested for IB's fformed
3 (independent) basic invariants.

The result is obtained through the computation of a metric matrix $\widehat
P(p)$, which is defined only in terms of the scalar products between the
gradients $\partial p_1(x),\dots \partial p_q(x)$, and whose domain of
semi-positivity is known to realize the orbit space $\real^n/G$ of $G$ as a
semi-algebraic variety in the space $\real^q$ spanned by the variables
$p_1,\dots ,p_q$.

After a short review of the approach that recently enabled to solve
the analogous problem for coregular groups with less than 5 basic
invariants, we determine the matrices $\wP(p)$ from the solutions of a
universal differential equation ({\em master equation}), which satisfy
new convenient additional conditions, which fit for the non-coregular
case.  The master equation involves as free
parameters only the degrees $d_a$ of the $p_a(x)$'s.  This approach bypasses
the actual impossibility of explicitely determining a set of basic polynomial
invariants for each group.

Our results may be relevant in physical contexts where the study of covariant
or invariant functions is important, like in the determination of patterns of
spontaneous symmetry breaking in quantum field theory, in the analysis of
phase spaces and structural phase transitions (Landau's theory), in covariant
bifurcation theory, in crystal field theory and so on.

\end{abstract}

\section{Introduction}

Invariant functions under the transformations of a
compact linear group (hereafter abbreviated in CLG) $G$, acting in an
Euclidean space $\real^n$, play an important role
in physics, and the determination of their properties is often a basic problem
to solve in many physical contexts, such as
the determination of patterns of spontaneous symmetry breaking, the
analysis of phase spaces and structural phase transitions (Landau's theory),
covariant bifurcation theory, crystal field theory and so on.

A $G$-invariant function $f(x)$, $\,x\in\real^n$, takes on constant values along
each orbit of $G$, thus, if one has to analyze its properties, it is certainly
more economical, and generally more effective, to think of it as a function
defined in the orbit space $\real^n/G$ of the action of $G$ in $\real^n$.  In
this way, it is possible to take fully into account the invariance properties
of $f(x)$, while maintaining its regularity properties, but avoiding the
troubles that could be met, for instance in the determination of the minima,
owing to its degeneracy along the $G$-orbits.

This approach to the study of the properties of a $G$-invariant function,
obviously requires a detailed knowledge of the structure of the orbit space
$\real^n/G$, which can be obtained as follows, using the methods of invariant
theory \cite{610}.  Let $\{p_1(x), \dots , p_q(x)\}$ be a minimal integrity
basis (hereafter abbreviated in MIB) of the ring $\real^n[x]^G$ of polynomial
invariants of $G$.  The MIB defines an {\it orbit map} $x\mapsto
\left( p_1(x),\ldots,p_q(x) \right)\, \equiv\, p(x)$,
mapping $\real^n$ onto a semi-algebraic subset, $p(\real^n)$, of $\real^q$,
which provides a diffeomorphic image of $\real^n/G$ (see, for instance,
\cite{610}, \cite{052} or \cite{710} and references therein).  If $f(x)$ is a
$G$-invariant polynomial or $C^\infty$-function it can be written in terms of
a polynomial or, respectively, $C^\infty$-function $\hat f(p)$,
in the form $f(x)=\hat f(p(x))$ \cite{285,617,690}.  The function $\hat f(p)$
has the same range as $f(x)$, but its domain is a faithful image of
$\real^n/G$.

The price to pay in the orbit space approach to the analysis of a
$G$-invariant function  is essentially twofold:

\begin{enumerate}
\item MIB's are sometimes difficult to determine.
\item The domain of the associated function $\hat f(p)$
is not the whole Euclidean space $\real^q$, but reduces to the semi-algebraic
subset $p(\real^n)$ \cite{020,021}, not trivial to determine.
\end{enumerate}

As for the problem stated under item 2, the following simple solution has been
proposed.  It has been shown \cite{020,021,651} that, if the relations among
the elements of the MIB's are known, the polynomial equalities and
inequalities defining $p(\real^n)$ and its strata can be determined from the
rank and positivity properties of a matrix $\wP(p)$, defined only in terms
of the gradients of the elements of a MIB.  When there are no relations among
the elements of the MIB's, that is for compact {\it coregular} linear
groups, the matrices $\wP(p)$ play the role of
inverse metric matrices, and the isomorphism classes of the orbit spaces of
all these groups can be classified in terms of equivalence classes of matrices
$\wP(p)$.

A way to obtain the matrices $\wP(p)$ generated by CLG's,
bypassing the actually insoluble problem of determining a MIB for each group,
has been suggested in \cite{681}.  The idea is to use an axiomatic approach,
that is to forget altogether the original definition of the matrices $\wP(p)$
and to characterize them as far as possible through their structural
properties ({\em allowable} \wPms).  The possibility of actually computing them
 is favored
by the identification and convenient formalization of a set of sufficiently
strong, but handy, analytic conditions, shared by all $\wP(p)$'s.  The less
immediately
evident of these conditions have been translated into a set of
differential equations ({\em boundary conditions}) involving the elements of
the matrices $\wP(p)$.

In the case of coregular groups, it has been proved that, in the case of less
than 5 basic invariants \cite{681,682,685}, the allowable $\wP$-matrices can
be determined from the solution of a unique universal differential equation
(that here we shall call {\em master equation}), satisfying convenient initial
conditions.  The master equation involves as free parameters only the degrees
$d_a$ of the basic polynomial invariants.  The conditions defining allowable
$\wP$-matrices for {\it coregular} CLG's turn out to be so restrictive
that, for each choice of a set of degrees $\{d_1,\dots , d_q\}$, they turn out
to select (at least for $q\le 4$) only a finite number of equivalence classes
of matrices $\wP(p)$.  Thus, even if the results obtained along these lines
are still strongly incomplete, one may reasonably hope to be able to obtain in
the future a classification of the orbit spaces of all compact coregular
linear groups,  even if the classification of these groups is not yet complete
and/or the explicit form of the elements of the
corresponding MIB's is not known.

In this paper,
we shall shortly review the geometry of linear group actions and
describe how
the invariant theory may successfully be applied to determine the
stratification of the orbit space of a compact linear group.  Then, we shall
turn our attention
on the axiomatic approach for the determination of the \wPms\/ of CLGs.
We shall focus on new developments concerning the \wPms\
generated by groups in the class ${\cal T}(q, q-1)$ of non-coregular groups
with only one
independent relation among the $q$ elements of their MIB's.
We shall show that the relation among the elements of a
MIB is determined by one of the irreducible factors of the determinant of
$\wP(p)$, which, along with $\wP(p)$, has to satisfy the master equation and
some additional conditions, part of which can be put in the form of a
subsidiary differential equation ({\em second order boundary conditions}).
The effectiveness of these conditions is tested in the simplest case of three
basic invariants, by determining all the solutions and selecting those which
lead to allowable \wPms.

In fact, the conclusion of our analysis is that, leaving aside the
trivial cases in which at least one of the invariants $p_1(x), p_2(x), p_3(x)$
is linear,

\begin{itemize}
\item There is only one monoparametric discrete family of allowable
non-equivalent
\wPms\ $\wP^{(k)}(p)$, $\natural\ni k\ge 2$, whose elements may be generated
by groups in ${\cal T}(3, 2)$.  The degrees of the $p_a$'s are
$d_1\,=\,d_2\,=\,k\,
\ge 2$, $d_3\,=\,2$ and, with a convenient choice of the $p_a$'s, the basic
relation can be written in the form $\widehat{F}^{(k)}(p)\,=\, p_1^2 + p_2^2 -
p_3^k $.

\item Every allowable \wPm\ of the family is generated by at least a
group $G\in {\cal T}(3, 2)$.

\item If the action of the groups is restricted to the unit sphere
$S^{(n-1)}$ of $\real^n$ (which is not essentially restrictive for what
concerns the characterization of the orbit space), all the orbit spaces
$S^{(n-1)}/G$, $G\in {\cal T}(3, 2)$ turn out to be isomorphic.
\end{itemize}

We shall present the matter in the following order.  In
Section 2 we shall
recall some known results concerning the characterization of orbit spaces
(see, for instance, \cite{683,710,080,021}) in the \wPm\ approach.
In Section
3 we recall the derivation of the boundary conditions, of the master relation
and of the additional conditions that must be satisfied in the case of
coregular CLG's.
Suitable conditions are subsequently derived for
non-coregular groups, with particular attention to the case in which there is
only one basic relation: this is the original part of the paper.
In \S 4 we formalize our approach to the
determination of the \wPms\ that could virtually be associated to actual
non-coregular CLG's with one independent relation among the basic invariants;
the boundary conditions and the master relation are presented as equations and
the notions of proper and allowable solutions of the master equation are
defined.  In \S 5 we compute first all the proper and allowable solutions of
the master equation and, after a further selection, we arrive at the
determination of the equivalence classes of the \wPms\ of all the non-coregular
CLG's of class ${\cal T}(3,2)$. The last part of the section is devoted to
the problem of the correspondence between allowable \wPms\ and  generating
CLG's.

Our conclusions agree with a result about locally smooth actions on
manifolds with orbits of codimension $\leq 2$
\cite[Th.~IV~8.2, p.~206]{080}. According to it, if the action of the
group is restricted to the unit sphere $S^{(2)}$ of $\real^3$, then
either $S^{(2)}/G$ is diffeomorphic to the unit interval $[0,1]$, or
$S^{(2)}/G$ is diffeomorphic to the $1$-sphere $S^{(1)}$.
The former case refers to coregular groups of class ${\cal T} (2,2)$,
the latter, both to non-coregular groups of class ${\cal T} (3,2)$
(whose orbit space is isomorphic to some $(\integer_m, \real^2)$, $m>1$,
where $\integer_m \subseteq \mbox{SO}(2,\real)$ acts on $\real^2$),
and to coregular groups of class ${\cal T}(3,3)$ (e.g. the
linear groups $\mbox{SO}(n,\real)$ acting in $\real^n \oplus \real^n$ for
$n \geq 3$).

The fact that coregular and non-coregular groups may share orbit spaces
belonging to the same isomorphism class is an intriguing fact that
suggested us to treat the axiomatic \wPm\/ approach stressing,
whenever possible, analogies and differences between those two cases.

It is worth noting that the effectiveness of our method does not
depend on the dimension of the real vector space upon which $G$ acts,
so it may be applied, in principle, to determine the orbit spaces of all
non-coregular groups with only one independent relation among the basic
polynomial invariants.

\section{An overview of the geometry of linear group actions}

In this section, we shall first define most of our notations and recall,
without proofs, some results concerning invariant theory and the geometry of
orbit spaces of CLG's, then we shall introduce the first definitions and the basic tools
for our subsequent analysis. For the unreferenced results see for instance
 \cite{080,710}.

\subsection{Orbits, strata and orbit spaces}

For our purposes, it will not be restrictive to assume that $G$ is a matrix
subgroup of $O_{n}(\real)$ acting linearly in the Euclidean space $\real^{n}$.

We shall denote by $x= (x_1,\dots ,x_n)$ a point of $\real^ n$. The group $G$
acts in $\real^n$ in the following way:

\eq{x'_i\,=\,(g\cdot x)_i\,=\,\summ 1nj g_{ij}\, x_j\,,\qquad x\in\real^n,\
g\in G\,.}
The $G$-orbit $\Omega_{x}$ through $ x\in\real^n$ and the {\em
isotropy subgroup} $G_{x}$ of $G$ at $x\in\real^n$ are
defined by the following relations:

\eq{\Omega_{x} \,=\, \{ g \cdot x \mid g \in G\},\qquad
G_{x} \,=\, \{g \in G \mid g \cdot x = x\}
\,.}

For all $x \in \real^n$, the isotropy subgroup $G_{x}$ at $x$
is a Lie group, which is not necessarily connected even if $G$ is.
If $G$ is continuous, the Lie algebra ${\cal G}_x$ of $G_{x}$ is formed
by the elements of the Lie algebra ${\cal G}$ of $G$ annihilating $x$;
moreover, the $G$-orbits are smooth, closed and compact submanifolds
of $\real^n$. They are connected if the group $G$ is.

For $G$ continuous, a tangent space
$T_{x}(\Omega)$ to an orbit $\Omega$ can be defined at all $x \in \Omega$.
It is formed by the tangent vectors at $x$ to regular curves through $x$,
lying in $\Omega$. Therefore $T_{x}(\Omega)$, as a vector space,
may be identified with $\{ a \cdot x, a \in {\cal G} \}$,
which is isomorphic to the Lie algebra quotient ${\cal G}/{\cal G}_{x}$.
The normal space $N_{x}$ to the orbit $\Omega$ through $x$ is the
orthogonal complement in $\real^n$ to $T_{x}(\Omega)$. It may be decomposed
into the direct sum $N_{x}^{(0)} \oplus N_{x}^{(1)}$, where $N_{x}^{(0)}$
denotes the orthogonal invariant space to $\Omega$ at $x$, formed by all
the vectors of $N_{x}(\Omega)$ which are invariant under $G_x$.
Since it can be prooved that each component space in the decomposition
$\real^n= T_{x}(\Omega) \oplus N_{x}^{(0)} \oplus N_{x}^{(1)}$ is globally
$G_x$-invariant, the representation $(G_x,\real^n)$ induced by $G$ in
$\real^n$ turns out to be completely reducible.

If $G$ is discrete, such as the finite subgroups of $\mbox{O}(n,\real)$, it
may be thought of as a compact Lie group with trivial connected component of
the unit.  The orbit $\Omega_{x}$ is a finite set formed by $\| G \|/ \| G_x
\|$ points.  Therefore, the tangent space at each point $x \in \real^{n}$
reduces to the null vector of $\real^n$ and the normal space is the entire
$\real^n$.

In any case, the invariance of the Euclidean norm under orthogonal
transformations assures that the G--orbit through $x$ is contained in the
sphere of radius $x$, centered in the origin of $\real^n$, while the linearity
of the action of $G$ in $\real^n$ implies

\eq{G_{x} \,=\, G_{\lambda x}, \,\, \forall \lambda \in
\real _*\,. }
The isotropy subgroup of $G$ at the origin of $\real^n$ coincides with $G$.
The isotropy subgroups of $G$, at points lying on the same orbit $\Omega_{x}$
are conjugate subgroups in $G$:

\begin{equation}
G_{g \cdot x} \,=\, g G_{x} g^{-1}, \, \forall g \in
G\,. \label{iso_group2}
\end{equation}

The class of all the subgroups of $G$ conjugate to $G_{x}$ in $G$ will be said
to be the {\em orbit type of} $\Omega_{x}$ and of its points; the orbit type
specifies the symmetry properties of $\Omega_{x}$ under transformations
induced by elements of $G$.

The points $x\in\real^n$ (or, equivalently, the orbits of $G$) sharing the
same orbit type form an {\it isotropy type stratum of the action of $G$ in}
$\real^ n$, hereafter called simply a {\it stratum of} $\real^n$.  All the
connected components of a stratum can be shown to be smooth iso-dimensional
sub-manifolds of $\real^n$.

Since any sufficiently small displacement from a point $x$ in the direction
orthogonal to $\Omega_{x}$ does not change the symmetry properties of $x$ iff
that direction belongs to $N_{x}^{(0)}$, it is possible to prove that the
following identity holds fot the tangent space $T_{x}(\Sigma)$ to the stratum
in $\real^n$:

\begin{equation}  \label {opli}
T_{x}(\Sigma) = T_{x}(\Omega) \oplus N_{x}^{(0)}\;.
\end{equation}

The {\em orbit space} of the action of $G$ in $\real^ {n}$ is defined as the
quotient space $\real^ n/G$ (obtained through the equivalence relation between
points belonging to the same orbit) endowed with the quotient topology and
differentiable structure.  We shall denote by $\pi$ the canonical projection
$\real^n\rightarrow\real^ {n}/G $.  Whole orbits of $G$ are mapped by $\pi$
into single points of $\real^{n}/G$.  Any function $f$ defined on $\real^n/G$
is differentiable iff $f \!\circ\!  \pi$ is differentiable on $\real^n$.  The
image through $\pi$ of a stratum of $\real^{n}$ will be called an ({\it
isotropy type}) {\it stratum of} $\real^{n}/G$; all its connected components
turn out to be smooth iso-dimensional manifolds.

Almost all the points of $\real^n/G$ belong to a unique stratum $\Sigma_p$,
the {\it principal stratum}, which is a connected open dense subset of
$\real^n/G$.  The boundary $\overline{\Sigma_p}\backslash\Sigma_p$ of
$\Sigma_p$ is the union of disjoint {\it singular} strata.  All the strata
lying on the boundary $\overline{\Sigma}\backslash\Sigma$ of a stratum
$\Sigma$ of $\real^n/G$ are open in $\overline{\Sigma}\backslash\Sigma$.

The following partial ordering can be introduced in the set of all the orbit
types:  $[H]<[K]$ if $H$ is conjugate to a subgroup of $K$.  The orbit type
$[H]$ of a stratum $\Sigma$ is contained in the orbit types $[H_b]$ of all the
strata $\Sigma_b$ lying in its boundary; therefore, more peripheral strata of
$\real^n/G$ are formed by orbits with higher symmetry under $G$
transformations.  The number of distinct orbit types of $G$ is finite and
there is a unique minimum orbit type, the {\it principal orbit type},
corresponding to the principal stratum; there is also a unique maximum orbit
type $[G]$, corresponding to the image through $\pi$ of the set of points of
$\real^n$, which are invariant under $G$; this set contains at least the
origin of $\real^n$.

Since any orbit $\Omega$ is mapped by $\pi$ into a single point of the orbit
space, \eref{opli} implies that the tangent space $T_{\Omega}(\hat \Sigma)$ to
a stratum $\hat \Sigma$ of $\real^n/G$ is isomorphic to the normal invariant
space $N_{x}^{(0)}$, where $x$ is any point belonging to the orbit $\Omega \in
\hat\Sigma$.  This fact has been exploited to construct a faithful image of
the orbit space $\real^n/G$ in a Euclidean space\cite{020,021}.  Before
reviewing this result, we shall recall a few basics of the geometric approach
to invariant theory.

A function $f(x)$ is said to be $G$-invariant if

\eq{f(g\cdot x)\,=\,f(x),\;\forall x\in \real^n,\ g\in G.}
The set of all real, $G$-invariant, polynomial functions of $x$ forms a ring
$\real [x]^G$, that admits a finite integrity basis \cite{285,617}.
Therefore, there exists a finite collection of invariant polynomials $p(x) = (
p_1(x), p_{2}(x), \ldots, p_{q}(x) )$ such that any element $F\in\real[x]^G$
can be expressed as a polynomial function $\widehat F$ of $p(x)$:

\eqll{F(x) \,=\, \widehat{F}(p(x)), \, \forall x \in \real^{n}\,.}{e2}
The polynomial function $\widehat F(p),\ p\in\real^q,$ will be said to have
weight $w$, if $w$ is the degree of the polynomial $F(x) = \widehat{F}(p(x))$
and it will be said to be $w$-homogeneous if $F(x)$ is homogeneous.  We shall
denote the homogeneity degree of $F$ by $w(F)$.

The elements of a basis of $\real[x]^G$ can be chosen to be homogeneous
polynomials.  The number $q$ of elements of a minimal integrity basis and
their homogeneity degrees $d_i$'s are only determined by the group $G$.

To avoid trivial situations, in this paper we shall only consider linear
groups with no fixed points, but for the origin of $\real^{n}$.  In this case,
the minimum degree of the elements of a minimal integrity basis is necessarily
2, and the following conventions can be adopted:

\begin{equation}
d_{1} \geq d_{2} \geq \ldots d_{q} \,=\, 2\,; \qquad
p_{q}(x) \,=\, \| x \|^2
 \,=\, \summ 1ni x_{i}^{2}\,.
\label{conventions}
\end{equation}
Hereafter, by a {\em minimal integrity basis} (abbreviated into {\em MIB}) we
shall always mean a {\it minimal homogeneous integrity basis} of the ring of
$G$-invariant polynomials, so chosen that the conventions of
\eref{conventions} hold true.

The orbits of a compact group $G$ are separated by the elements of any MIB of
$G$, i.e., at least one element of a MIB takes on different values on two
distinct orbits.  But it can be said more.  Each MIB $\{p_1(x),\dots
,p_q(x)\}$, defines an orbit map $p:\, \real^{n} \longrightarrow \real^{q}$,
$x \mapsto \left( p_1(x),\ldots,p_q(x) \right)$, which maps all the points of
$\real^{n}$ lying on an orbit of $G$ into a single point of $\real^{q}$.  The
range $p(\real^n)$ of the orbit map $p$ yields a faithful image of the orbit
space of $G$, and the elements $\{p_1, \dots,p_q\}$ of the MIB, thought of as
coordinates in the space $\real^q$, provide a smooth parametrization of the
points of $\real^n/G$.  In fact, it can be shown that every orbit map induces
a diffeomorphism of $\real^n/G$ onto a semi-algebraic connected closed subset
$\overline{{\cal S}}$ of $\real^q$:  \eqll{ \overline{{\cal S}} \,=\,
p(\real^{n}) \, \simeq \, \real^n/G\;.
} {essebar}

The analysis of the structure of $\real^n/G$ is easier if one confines his
attention to the orbit space of the action of $G$ on the unit sphere $S^{n-1}$
of $\real^n$.  This is not restrictive for the following reasons.  Owing to
the linearity of the action of $G$, the isotropy subgroups of $G$ at points
lying on the same straight line through the origin of $\real^n$ coincide; thus
an essentially complete specification of the structure of $\real^n/G$ is
obtained from the structure of $S^{n-1}/G$.  Indeed, there is a bijection
$\Phi$ mapping the set $\{\Sigma_1\}$ of strata of $S^{n-1}/G$ onto the set
$\{\Sigma\}$ of strata of $(\real^n\backslash \{0\})/G$, such that
$\Sigma=\Phi (\Sigma_1 )$ is homeomorphic to $\Sigma_1\times \real_+$.
Moreover, the orbit space $S^{n-1}/G$ is compact and connected.  The same is
consequently true for its image under an orbit map $p$:  the semi-algebraic
set \eqll{ {\overline{\cal S}}_1=p(S^{n-1})={\overline{\cal S}}\cap \Pi,\qquad
\Pi =\{p\in\real^ q\mid p^q=1\}}{115} is compact and connected.

\subsection{Coregular and non--coregular groups}

A set $\{p_1(x),\dots ,p_q(x)\}$ of $G$-invariant polynomials will be said to
be {\em regular} if its elements are algebraically, and therefore functionally,
independent.  The linear group $G$ will be said to be {\em coregular} if its
ring of invariant polynomials admits regular integrity bases.

If $\{p\}$ is a MIB of a coregular linear group, the polynomial function
$\widehat F(p)$ appearing in Eq. (\ref{e2}) is uniquely determined and, when it
is $w$-homogeneous it satisfies the following relation, which is an immediate
consequence of Euler equation $\summ 1ni x_i\partial_i F(x) = w(F) F(x)$:

\eqll{\summ 1qa d_a p_a \, \partial_{a} \widehat F(p) =
w(F)\widehat F(p).}{Euler}

Let $\{\widehat F_A(p)\}_{1\le A\le K}$
be a complete set of basic homogeneous relations among the elements of a
non-regular set $\{p_1(x),\dots ,p_q(x)\}$ of $G$-invariant homogeneous
polynomials.  The polynomials $\widehat F_A(p)$ can be chosen to be
$w$-homogeneous and irreducible on the complex numbers.  The associated
equations

\eqll{ \widehat F_A(p)\,=\,0,\qquad A=1,\dots, K}{e5}
define an irreducible algebraic variety in $\real^q$
(and in $\complex^{q}$ for $ p \in \complex^{q}$), which will be called the
{\em variety $\cal Z$ of the relations} among the elements of the set. The
variety $\cal Z$ has a singularity in $p = 0$.
In fact, for all $A$, $\widehat F_A (p)$
is a $w$-homogeneous polynomial which cannot be solved polynomially
with respect to anyone of the basic invariants $p_a$.
The absence of linear terms in any $p_a$ implies:
\eqll{\widehat{F}_{A}(0)\, =\, 0 \,; \qquad
 \partial\widehat{F}_{A}(0)\ =\,0,\qquad A=1,\dots, K\,.}{sing}
%%%%%%
For $k=\mbox{dim}({\cal Z})$, the couple
$( q ,k)$ will define the {\it regularity
type} (hereafter called $r$-type) of the set $\{ p \}$.
If $\{p\}$ is a MIB of a group $G$, the couple
$(q,k)$ will define the {\it $r$-type} of $G$.
If
$G$ is coregular, there are no relations among the elements of its MIB's and
${\cal Z}=\real^q$.

It will be worthwhile to note that, if there are relations among the elements
of an integrity basis, the polynomial function $\widehat F(p)$ appearing in
(\ref{e2}) is uniquely determined on $\cal Z$, but, as a polynomial function
on $\real^q$, it is only determined modulo a polynomial $\widehat F_0(p)$
vanishing identically on $\cal Z$.  In any case, however, its weight is
uniquely determined and it will always be possible to choose $\widehat F(p)$
so that it satisfies (\ref{Euler}).  In the following, by a {\em
$w$-homogeneous polynomial in $p$} of weight $w$ we shall always mean a
polynomial satisfying (\ref{Euler}).

\subsection{\label{P12}The $\widehat{P}(p)$ matrix}
Let $\{p(x)\} = \{p_1(x),\dots ,p_q(x)\}$ be a MIB for the
group $G$ and  $d_1,\dots
,d_q$, the corresponding weights.
We shall associate to $\{p(x)\}$ the following square matrix, whose
elements are $G$-invariant polynomials in $x$, defined only in terms of the
$G$-invariant Euclidean scalar products $\scalarp \cdot\cdot$ between the
gradients of the elements of the set $\{p(x)\}$:

\begin{equation}
P_{ab}(x) \,=\, \scalarp {\partial p_a(x)}{\partial p_a(x)}\, =\,
\summ 1ni\frac{\partial p_a(x)}{\partial x_i}\cdot\frac{
\partial p_b(x)}{\partial x_i} \,=\, \widehat{P}_{ab}(p(x)) \,, \label{matP}
\end{equation}
where $a,b = 1, \ldots, q$.
In the last member of (\ref{matP}), use has been made of Hilbert's theorem, in
order to express $P_{ab}(x)$ in terms of polynomial functions $\widehat
P_{ab}(p_1, \dots ,p_q)$.  As already noted, these functions
are uniquely determined only on $\cal Z$.

The following properties, which are common to all the matrices $\widehat
P(p)$, are more or less immediate consequences of their definition
and of the conventions we have adopted:

 \vskip2truemm

\noindent{\bf P1.} {\em Symmetry, homogeneity and bounds on the last row and
column:}\ The matrix $\widehat{P}(p)$ is a real $q \times q$ symmetric matrix,
whose elements $\widehat{P}_{ab}(p)$ can be chosen
to be $w$-homogeneous polynomials of weight

\eqll{w(\widehat{P}_{ab}) \,=\, d_{a} + d_{b} - 2\,.}{weightP}
Owing to the definition $p_q(x) = \summ 1qi x_i^2$, the last row and column
of every matrix $\widehat {P}(p)$ are determined by the degrees of the MIB:

\eqll{ \widehat{P}_{qa}(p) \,=\, \widehat{P}_{aq}(p) \,=\, 2 d_{a} p_{a},
\qquad a=1, 2, \ldots, q\,.}{LastRow}
%%%%%%%%%%%%%
 \vskip2truemm
The orbit space $\overline{{\cal S}}$,
is a
connected semi-algebraic sub-variety of $\real^{q}$ and, like all
semi-algebraic varieties \cite{840}, it presents a natural stratification in
connected semi-algebraic sub-varieties ${\sigma_{i}}^{(\alpha)}$, called {\em primary
strata}\footnote{A simple example of a compact connected  semi-algebraic
variety of $\real^3$ is yielded by a polyhedron.  Its interior points form its
unique 3-dimensional primary stratum, while 2-,1- and 0-dimensional primary
strata are formed, respectively, by the interior points of each face, by the
interior points of each edge and by each vertex.}.
%%%%%%%%%%%%%
The set $\overline{{\cal S}}$, therefore, consists in a finite collection of disjoint
and connected semi-algebraic submanifolds of $\real^{q}$, $\,\{
{\sigma_{i}}^{(\alpha)} \}_{i,\, \alpha}$, such that $\overline{{\cal S}} =
\bigcup_{i\,, \alpha}{\sigma_{i}}^{(\alpha)}$ and the boundary of each
${\sigma_{i}}^{(\alpha)}$ is vacuum, or the union of lower-dimensional
${\sigma_{j}}^{(\beta)}$'s, $(j>i)$. Each ${\sigma_{i}}^{(\alpha)}$ is
open in its closure and is defined recursively on $i$
(which distinguishes semi-algebraic sets of different dimensions)
 as the $\alpha$-th
connected component of the set of regular points of the semi-algebraic
varieties ${\cal W}_{i} = \overline{{\cal S}} \setminus
\bigcup_{0\, <j< \,i}{\sigma_{j}}^{(\beta)}\,,\;\;i=1,2\ldots .$

%%%%%%%%%%

\vskip2truemm

A characterization of the image $\overline{{\cal S}}\,=\,p(\real^n)$
of the orbit space of $G$ as a
semi-algebraic variety can be easily obtained through the matrix
$\widehat{P}(p)$ associated to one of its MIB's.

Let us introduce the
result  that permits to get an advantage
out of the
analysis in \S 2.1 about the local properties of the action of
$(G,\real^n)$. It was prooved in \cite{680} that
the normal invariant space $N_{x}^{(0)}$ through $x$ coincides with the
vector space generated by the set
$\Delta=\{ \partial p_l(x) \}_{1 \leq l \leq q}  $
made up of the gradients at $x$ of the elements of a MIB.
The $\wP(p)$ matrix is then defined from the grammian matrix associated
to the set of vectors $\Delta$, which
are a basis for the tangent space to the stratum $\hat \Sigma$ containing
an orbit $\Omega$ through $x\in \real^n$.
Therefore, the following fundamental
theorem \cite{020,021,651} holds true:

\begin{theorem}  \label{T1}
Let $G$ a compact matrix subgroup of $\mbox{O}_{n}(I\!\!R)$, $p$ the orbit map
$\real^n\rightarrow\real^q$ defined by the homogeneous MIB
$\{p_1(x),p_2(x),\ldots ,p_q(x)\}$ and $\widehat{P}(p)$ the matrix defined in
{\rm \eref{matP}}.  Then $\overline{{\cal S}}= p(\real^{n})$ is the
 semi-algebraic subset of the variety ${\cal Z}\subseteq \real^q$ of
the relations among the elements of the MIB where $\widehat{P}(p)$ is positive
semi-definite.  The set $\overline{{\cal S}}$ is connected.
The $k$--dimensional primary strata of $\overline{{\cal S}}$
are the connected components of the set $\widehat{W}^{(k)} = \{ p \in {\cal
Z}\; \mid \widehat{P}(p) \geq 0, \, \mbox{{\rm rank}} (\widehat P(p)) = k \}$; they
are the images of the connected components of the k--dimensional isotropy type
strata of $\real^{n}/G$.  In particular the interior ${\cal S}$ of
$\overline{{\cal S}}$, where $\widehat{P}(p)$ has the maximum rank,
is the image of the principal stratum and is connected.
\end{theorem}

The theorem assures that the orbit space of a coregular group is completely
determined by the positivity conditions of a $\widehat P$-matrix computed from
any one of its MIB's. For non-coregular groups, also a complete set of
relations among the $p_a$'s has to be specified; this however can be
obtained from rank conditions on the matrix $\wP(p)$.

\vskip3truemm
We shall also need the following property of the $\widehat P$-matrices
associated to MIB's of CLG's:

\vskip2truemm
\noindent{\bf P2. }{\em Tensor character:} If $\{p_1,\dots ,p_q\}$ is a
MIB, the matrix elements of $\widehat{P}(p)$
transform as the components of a rank 2 contravariant tensor under MIB
transformations that maintain the conventions fixed in \eref{conventions}
(these transformations will be hereafter called MIBT's).  In fact, let $\{
p(x)\}$ and $\{p'(x)\}$ be distinct MIB's; the $p'_a(x)$'s, being G--invariant
polynomials, can be expressed as polynomial functions of the
$p_a(x)$'s\footnote{Since in our conventions the $q$-th element
of any MIB is fixed, when defining a MIBT we shall
always understand the condition $p'_q=p_q$.}:

\begin{equation} \label{MIBT}
\begin{array}{rcl}
{p'}_{\alpha}  &=&  {p'}_{\alpha}(p), \qquad\alpha = 1, \ldots, q - 1\,,  \\
&&\\
p'_q &=& p_q\,,
\end{array}
\end{equation}
where each of the polynomial functions $p'_\alpha(p)$ depends only on the
$p_{\beta}$'s whose weights $d_{\beta}$ are not greater than $d'_\alpha $. Then,

\begin{equation} \label{equiv}
\widehat{P}'(p'(p)) \,=\, J(p) \cdot \widehat{P}(p) \cdot J^{T}(p) \,,
\end{equation}
where we have denoted by $J(p)$ the Jacobian matrix of the transformation:

\eq{J_{ab}(p) \,=\, \partial {p'}_{a}(p)/ \partial p_{b},\qquad a,b=1,\dots
,q\,;}
the matrix $J$ turns out to be upper--block triangular, with constant
elements in the diagonal blocks, so that the determinant of $\widehat{P}(p)$
is a relative invariant of the group of the MIBT's.

\subsection{Classification of the orbit spaces of compact linear groups
\label{Cosclg}}
Two \wPms\ $\wP(p)$ and $\wP'(p')$, computed from different MIB's $\{p\}$ and
$\{p'\}$ of
the same CLG will be said to be {\em equivalent}; the semi-algebraic
varieties $\overline{{\cal
S}}$ and $\overline{{\cal S'}}$ defined by the positivity conditions imposed
on $\widehat{P}(p)$ and $\widehat P'(p')$ respectively, are isomorphic (as
semi-algebraic varieties) realizations of the orbit space $\real^{n}/G$.

As stressed in the introduction, we shall try to characterize the
\wPms\ through their structural properties.
So we shall need to extend the notion of equivalence to matrices endowed
with the formal properties of the \wPms.

Let $\{p_1,\dots ,p_q\}$ and $\{p'_1,\dots ,p'_q\}$ be two sets of ''weighted"
indeterminates, sharing the same set of
weights $\{d_1,\dots ,d_q\}$, satisfying the first of our conventions in
(\ref{conventions}).  Two matrices $\widehat{P}(p)$ and $\widehat P'(p')$,
satisfying the conditions stated under {\bf P1} will be said to be {\em equivalent}
if they are connected by a relation like (\ref{equiv}), where $J(p)$ is the
Jacobian matrix of the transformation $p'= p'(p)$ endowed with the same formal
properties of a MIBT.

The semi-algebraic varieties where two equivalent
matrices $\widehat{P}(p)$ and $\widehat P'(p')$ turn out to be positive
semi-definite, are clearly isomorphic.

On the basis of what we have just said, the notions of MIBT's
(\ref{MIBT}) and of equivalence of $\widehat{P}$-matrices  \eref{equiv}
can be extended to the case of different groups $G$ and $G'$,
under suitable conditions.

\begin{definition}
Let $\{p_1,\dots ,p_q\}$ and $\{p'_1,\dots ,p'_q\}$ be MIB's respectively of
the compact linear groups $G$ and $G'$, sharing the same degrees $d'_j=
d_j,\ j=1,\dots,q$. The orbit
spaces $\real^{n}/G$ and $\real^{n'}/G'$ will be said to be isomorphic if
there exists a formal MIBT $p'=p'(p)$ such that:
\begin{itemize}
\item[i)] for every $\widehat F'(p')\in {\cal I}({\cal Z}')$, the
function $\widehat{F}(p)=\widehat F'(p'(p)) \in {\cal I}({\cal Z})$;
\item[ii)] the $\widehat P$-matrices $\widehat{P}(p)$ and $ \widehat P'(p')$,
associated to $\{p\}$ and $\{p'\}$ are equivalent.
\end{itemize}
\end{definition}

If $G$ and $G'$ have isomorphic orbit spaces, then the images of their orbit
spaces $\overline{{\cal S}}$ and $\overline{{\cal S'}}$, associated with the
MIB's $\{p\}$ and $\{p'\}$ are isomorphic semi-algebraic varieties:

\eq{\overline{{\cal S'}} \,=\, p'(\overline{{\cal S}})\,.}

Thus, in particular, the classification of the isomorphism classes of the
orbit spaces of the {\em coregular} CLG's rests on the determination of a
representative for each class of equivalent $\widehat{P}(p)$ matrices.
As already noted, the orbit space of a non--coregular group can be determined
from the knowledge of the \wPm\ associated to one of its MIB's only if a
complete set of basic relations among the elements of the MIB is specified.

\section{Characterizing the matrices $\wP(p)$}
In this section we shall point out a set of additional conditions that
should characterize, as far as possible, the \wPms\ associated to CLG's.

\subsection{\label{P3}Boundary conditions}
It has been shown in \cite{681,683} that,
besides the constraints listed in \S \ref{P12} under {\bf P1},
 every \wPm\ has to satisfy some additional
conditions, that can be put in the form of a set of differential relations,
so that
one can try to determine the associated  \wPms\ associated
to CLG's as solutions of a system of differential
equations.  Let us briefly recall the derivation of these results.

%%%%%%%%%%%%%%%
Let us denote by $\sigma$ a general primary stratum of $\overline{{\cal S}}$,
and  by ${\cal
I}({\sigma})$ the ideal formed by all the polynomials in $p \in \real^{q}$
vanishing on ${\sigma}$.  Every $\hat{f}(p)\in {\cal I}({\sigma})$
defines in $\real^{n}$ an invariant polynomial function
$f(x) = \hat{f}(p(x))$, and
\begin{equation}
f(x)  \,=\, 0, \hspace{3em} \forall x \in \Sigma_{f}=p^{ -1}({\sigma})\,.
\end{equation}
The gradient $\partial f(x)$ is obviously orthogonal to $\Sigma_f$ at
every $x\in\Sigma_f$, but, it must also be tangent to $\Sigma_{f}$
since $f(x)$ is a G--invariant function \cite{080,680}. As a consequence, it
has to vanish on $\Sigma_{f}$:

\begin{equation} \label{A1}
0\,=\,\partial f(x) \,=\, \left.\summ 1qb\partial_b \hat{f}(p)
\, \partial p_{b}(x) \right|_{p=p(x)}
\,, \hspace{2em} \forall x \in \Sigma_{f} \,.
\end{equation}
By taking the scalar product of (\ref{A1}) with $\partial p_{a}(x)$, we
end up with the following {\em boundary conditions}:

\begin{equation} \label{A2}
\summ 1qb \widehat{P}_{ab}(p) \, \partial_b \hat{f}(p) \in
{\cal I}({\sigma}), \hspace{2em} \forall \hat{f} \in {\cal I}
({\sigma}) \,\, \mbox{and}
\,\, \forall {\sigma} \subseteq \overline{{\cal S}}\,.
\end{equation}
Equation (\ref{A2}) can be re-proposed in the form of a differential relation
involving only polynomial functions of $p$.
%%%%%%%%%%%%%%%
According to the Hilbert basis theorem \cite{Ken}, the ideal ${\cal
I}(\sigma)$ is finitely generated.
Let $\{ f^{(1)}(p), f^{(2)}(p), \ldots, f^{(m)}(p) \}$ be a $w$-homogeneous
basis for ${\cal I}({\sigma})$, then \eref{A2} is equivalent to the
following relations:

\begin{equation} \label{A3}
\summ 1qb \widehat{P}_{ab}(p) \, \partial_b a^{(r)}(p)
 \,=\, \summ 1ms \lambda^{(rs)}_a(p) \, a^{(s)}\,,
\qquad a=1,\ldots,q\,; \;\; r=1,\ldots,m \,,
\end{equation}
where the $\lambda^{(rs)}$'s are $w$-homogeneous polynomial functions of $p$
of weight $(w(a^{r}) - w(a^s) + d_a -2)$.

It is easy to realize that
\vskip3truemm
\noindent{\bf P3.}{\em The $a^{(r)}$ transform like relative
invariants and the $\lambda^{(rs)}(p)$ like vector fields, under MIB
transformations.}
\vskip3truemm
In the particular case in which $\sigma$ is a $(q-1)$-dimensional primary
stratum, the ideal ${\cal I}(\sigma)$ has a unique {\it irreducible}
generator, $a(p)$, and \eref{A3} reduces to the simpler form
%\cite{681,682,683}

\begin{equation} \label{mas}
\summ 1qb \widehat P_{ab}(p) \, \partial_b a(p) \,=\, \lambda_a(p)\, a(p)\,,
\qquad a=1, \dots, q\,.
\end{equation}
Equation (\ref{mas}) will be quoted as {\em master relation}\footnote{
The present extension of the \wPm\ approach to the non-coregular case
has stimulated a refinement
of the definition of the {\em canonical equation} with respect to the
one appeared in \cite{681,682,683}.}.

There are only two types of $(q-1)$-dimensional strata:

\begin{enumerate}
\item Sub-principal strata of orbit spaces of coregular groups.
 This case will be discussed in \S~\ref{AddC}.

\item Principal strata of orbit spaces of groups
of $r$-type $(q,q-1)$. This case will be discussed in \S~\ref{AddNC}.
\end{enumerate}

The structure of  (\ref{mas}) has been analyzed in Ref. \cite{682}
where the results summarized below have been proved.
We shall need them in the following.

\begin{itemize}
\item[i)]
$a(p)$ is a polynomial factor of $\det \wP(p)$; it will be called
an {\it active} factor of $\det\wP(p)$.

\item[ii)] If, for a given $\widehat P(p)$, the couples $(a^{(i)}(p),
\lambda^{(i)}(p))$, $i=1,\dots ,K$, satisfy the master relation (\ref{mas}),
then the couple $(A(p) = \prodd 1Ki a^{(i)}(p),\;\lambda(p) =\summ 1Ki
\lambda^{(i)}(p))$ satisfies the master relation.

\item[iii)] If, for a given MIB $\{p'\}$, the couple $(A'(p'), \lambda'(p'))$,
satisfies the master relation (\ref{mas}), there exist particular MIB's, which we shall
call {\it $A$-bases}, in which the vector $\lambda (p)$ reduces to
the simple {\it canonical} form:

\eqll{ \lambda_a(p)=2\delta _{aq}w(A)\, , \qquad a=1,\ldots,q\,.}{sqx1}

In an $A$-basis, the master relation assumes the following {\it
canonical} form:

\eqll{\summ 1qb \widehat P_{ab}(p)\partial_b A(p) \,=\, 2\delta_{aq}
w(A) A(p),\qquad a=1,\dots ,q\,.}{can}
\end{itemize}

\begin{remark}
{\em
The studies about structural phase transitions in the Landau approach have
stimulated some authors \cite{Jaric,454,455} to examine the problem of the
determination of a basis for covariant vector fields (or, in general, tensor
fields) for the action of a (point) group $G$.  Although the solution of this
problem is not necessary to determine the minima of the Landau free energy,
the role of the vector fields was analized in detail in the mathematical
literature \cite{052,710}.  In particular, it has been prooved that a smooth
vector fields $V$ on the image of the orbit space $\overline{\cal S}$ is
tangent to a stratum $\sigma$ iff it preserves the ideal ${\cal I}(\sigma)$ in
${\cal C}^{\infty}(\overline{\cal S},\real)$ of the real valued smooth
functions vanishing on $\sigma$.

In this context, the columns of the \wPm\/ may be viewed as the components
of a strata preserving vector field. }

\end{remark}

%%%%%%%%%%%%%%%%%%%%%%%%%%%%%%%%%%%%%%%%%%%%%%%%%%%%%%%%%%%%%%%%%%%%%%%%%
\subsection{\label{AddC}Additional conditions for coregular groups}

If $\{p\}$ is a regular MIB, there is a unique (generally reducible)
generator, $A(p)$, of the ideal ${\cal I} ({\cal B})$, associated to the union
$\cal B$ of all the $(q-1)$-dimensional strata of $\overline{\cal
S}$\footnote{The closure of $\cal B$ forms the boundary of $\overline{\cal
S}$.} and it satisfies (\ref{can}).

The following results have been proved in \cite{682} to hold true in every
$A$-basis:

\vskip3truemm
\begin{itemize}
\item[i)] The point $p^{(0)}\,=\,(0,\dots , 0,1)$ lies in the interior $\cal
S$ of $\overline{\cal S}$; it is the image of a particular $G$-orbit lying on
the unit sphere of $\real^n$.

\item[ii)] $A(p)$ is a factor of $\det \widehat P(p)$; it can be normalized
at $p^{(0)}$:

\eqll{A(p^{(0)})\,=\,1}{a2b}

and its weight is bounded:

\eqll{2d_1\le w(A)\le w(\det\widehat P)\,=\,2\summ 1qa d_a -2q\,.}{a2a}

\item[iii)] The restriction $A(p)\big |_{p_q=1}$, of $A(p)$ to the hyperplane
$\Pi$ of $\real^q$, has a unique local non
degenerate maximum lying at $p^{(0)}$; thus:

\eqll{\left .\partial_\alpha A(p)\right|_{p=p^{(0)}}\,=\,0,\qquad \alpha =1,
\dots, q-1\,.}{gradA}

\item[iv)] $\widehat P(p^{(0)})$ is block diagonal, each block being
associated to a subset of $p_a$'s sharing the same weight, and, in a subclass
of $A$-bases ({\it standard $A$-bases}), it is diagonal:

\eqll{\widehat P_{ab}(p^{(0)})\,=\,d_ad_b\delta_{ab}, \qquad a,b=1, \dots,
q.}{iniP}

% Two different standard $A$-bases are related by a MIBT not involving $p_q$:
%
% \eqll{p'_\alpha \,=\,f_\alpha (p_1,\dots ,p_{q-1}),\qquad \alpha = 1, \dots ,
% q-1\,;}{a2d}
%the corresponding Jacobian matrix is orthogonal at $p^{(0)}$.

\end{itemize}

\subsection{\label{AddNC}Additional conditions for non-coregular groups}
The \wPms\ generated by non-coregular groups do not satisfy the set of
``initial conditions'' specified in the preceding subsection, but the
presence of relations connecting the elements of any MIB gives rise to
constraints that we shall try and formalize in a convenient way in this
section.
We shall first deal with
general non-coregular groups;
subsequently the results will be specialized to groups
of $r$-type $(q,q-1)$.

\subsubsection{\label{Sobcgc} Second order boundary conditions.
The general case}

Let us consider a compact non-coregular group $G$, whose orbit space,
according to Theorem~\ref{T1}, is realized as a
semi-algebraic subset $\overline{{\cal S}}$ of the variety ${\cal Z}$ of the relations.

We shall denote by ${\cal I}({\cal Z})$ the ideal of the polynomial
functions of $p$ vanishing on $\cal Z$.  Any polynomial $\widehat{F}(p) \in
{\cal I}({\cal Z})$ defines an identity in $\real^{n}$:

\begin{equation} \label{BB1}
F(x) \,=\, \widehat{F}(p(x)) \,=\, 0 \,,
\end{equation}
which, after differentiating twice with respect to $x_i$
and summing over $i$, gives rise to
the following condition, valid $\forall x \in \real^{n}$:

\begin{equation} \label{BB3}
 \summ 1ni \left \{ \summ 1q{a,b}
\left( \partial_{a} \partial_{b} \widehat{F}\right)\!(p(x))
\, \partial_{i} p_{b}(x) \,
   \partial_{i} p_{a}(x) +
\summ 1qa \left( \partial_{a} \widehat{F}\right)\!(p(x))
\, \partial^{2}_{i} p_{a}(x)
\right \}  \,=\, 0 \,.
\end{equation}

Since $G$ is a matrix subgroup of $\mbox{O}_{n}(\real )$, the $n$-dimensional
Laplacian of any invariant polynomial function of $x$ is a
$G$-invariant polynomial.  Thus Hilbert's theorem ensures the existence of a
collection of polynomial functions $\hat l_{a}(p) \,\, (a=1, 2, \ldots, q)$ of
$\real^{q}$ such that:

\begin{equation}\label{BB3'}
l_a(x)\ =\  \summ 1ni \partial^{2}_{i} p_{a}(x) \,=\,
\hat l_{a}(p(x))\,, \qquad a=1, 2, \dots, q \,.
\end{equation}

From their very definition the $l_a(x)$'s are
homogeneous polynomials with degrees $d_a - 2$;
therefore the $\hat l_a(p)$ can be chosen to be $w$-homogeneous
polynomials in $p$ with weights
\eqll{ w(\hat l_{a}) = d_{a} - 2.}{BB3xy}
Since in our approach the explicit form of the polynomials
$ p_{\alpha}(x), \; \alpha=1,2,\ldots,q-1\,,$ is not specified, in the
following the $\hat l_{\alpha}(p)$
have to be thought of as unknown $w-$homogeneous polynomials, while,
owing to the
convention $p_{q}=\sum_{j=1}^{n} x_{j}^{2}$:

\begin{equation}\label{BB3''}
l_{q} \,=\, 2 n\,\ge 4\,
\end{equation}
and the inequality is a consequence of the fact that, in our assumptions,
the dimension $n$ of the vector space in
which $G$ acts must be $\ge 2$.
%%%%%%%%%%%%%%%%%%%%%%%%%%%%%%

The identity expressed in (\ref{BB3}) induces, through the orbit map, the
following polynomial relation, valid for all $p\in \overline{{\cal S}}
\,=\,p(\real^{n})$,
and consequently, for all $p \in  {\cal Z}$,
as $p(\real^{n})$ is a semi-algebraic subset of $\cal Z$ of the same
dimension as $\cal Z$:

\begin{equation} \label{FE1}
\summ 1q{a,b} \widehat{P}_{ab}(p)\;\partial_{a}\partial_{b} \widehat{F}(p)
  +
 \summ 1qa \hat{l}_{a}(p)\; \partial_{a} \widehat{F}(p)
     \,=\, 0 \,, \qquad   p \in {\cal Z}\;.
\end{equation}

It is now evident that (\ref{FE1}) may be considered as a sort of {\em
second order boundary condition.}
%%%%%%%%%%%%%%%%%%%%%%

Let us denote by $\{\widehat{F}_{A}(p)\} , \;$ $1\le A\le
K$ a $w$-homogeneous minimal basis of the polynomial ideal
${\cal I}({\cal Z})$, then (\ref{FE1}) can be
rewritten in the following equivalent form:

\begin{equation} \label{FE2}
\summ 1q{a,b} \widehat{P}_{ab}(p) \;
\partial_{a} \partial_{b} \widehat{F}(p)\:
   +
\summ 1qa \hat l_{a}(p) \,\; \partial_{a} \widehat{F}(p)
  \,=\, \summ 1mA \xi_{A}(p) \widehat{F}_{A}(p)\,\,,
\end{equation}
where the $\xi_{A}(p)$'s are $w$-homogeneous polynomials.

The r.h.s. of (\ref{FE2}) has to be a $w$-homogeneous polynomial of
weight $(w(\widehat{F}) - 2 )$ like the l.h.s..
As a consequence, in the particular case in which $\widehat{F}$ is a
lowest weight element of the ideal ${\cal I}({\cal Z})$, the second member
of Eq.~(\ref{FE2}) must be zero.  Thus we have proved
the following proposition:

\begin{proposition} \label{HP}
Let $\widehat{F}(p)$ be a lowest weight generator of the ideal ${\cal
I}({\cal Z})$ generated by the polynomial relations among the elements of a
MIB.  Then

\begin{equation} \label{FE3}
\mbox{\rm Trace} \, \left(\widehat{P}(p) \cdot
\mbox{\rm He}\left ( \widehat{F}(p) \right )
\right ) + \scalarp{\hat l(p)}{\partial\widehat F(p)} \,=\,0 \,,
\end{equation}
where
$\scalarp\cdot\cdot$ denotes the Euclidean scalar product in $\real^q$,

\begin{equation} \label{FE3'}
\left ( \mbox{\rm He}\left ( \widehat{F}(p) \right )\right)_{a,\,b} \,=\,
\partial_a\partial_b \widehat{F}(p)\,,\qquad a,b=1,\dots ,q\,,
\end{equation}
and the  $\hat l_a$'s have been defined in {\rm (\ref{BB3'})}.
\end{proposition}

\subsubsection{Groups of r-type $(q,q-1)$}

In the case of groups of r-type $(q,q-1)$, the ideal ${\cal I}({\cal
Z})$ of the relations among the elements of a MIB $\{p\}$ has a unique
generator $\widehat F(p)$, so that the master relation is satisfied by
$\widehat F(p)$:

\begin{equation} \label{mas1}
\summ 1qb \widehat P_{ab}(p) \, \partial_b \widehat F(p)
 \,=\, \lambda_a(p)\, \widehat F(p)
\end{equation}
and Eq.~(\ref{FE3}) holds also true.

The second order derivatives of $\widehat F(p)$ in (\ref{FE3})
can be eliminated.  In fact, by
differentiating (\ref{mas1})
with respect to $p_a$ and summing over $a$, one obtains

$$\summ 1q{a,b} \wP _{ab}(p)\partial_a\partial_b\widehat F(p)
\,=\, \summ 1qb\left(-\summ 1qa \partial_a\wP _{ab}(p) +
\lambda_b(p)\right)\partial_b\widehat F (p)\,,$$
where use has been made of the fact that  $\partial \lambda_a/\partial p_a
= 0$, since
the weight of $\lambda_a$ is $d_a - 2$.
After replacing in (\ref{FE3}) and
defining the $w$-homogeneous vector field $L_a(p)$ by
\eqll{L_a(p)\,=\, \hat l_a(p) + \lambda_a(p) - \summ 1qb \partial_b\wP _{ab}(p)
\,,}{defL}
we obtain
\eqll{\summ 1qa L_a(p)\partial_a\widehat F(p) \,=\, 0\,.}{S1}

The weights of the components of $L(p)$ are the following:

\eqll{w(L_a) \,=\, d_a - 2.}{weightL}
In particular $L_q$ is a constant that can be easily calculated,
from (\ref{LastRow}), (\ref{sqx1}) and (\ref{BB3''}):

\eqll{L_q = 2\left(n + w(F)  - \summ 1qa d_a\right)\, =\, 2D\,,
}{Lq}
with
\eqll{D\,\ge\,w(F) - \summ 1{q-1}\alpha d_\alpha\,,}{S2}
where the lower bound is a consequence of the relations
$d_q = 2$ and $n\ge 2$.

\section{The master equation and allowable \wPms}
%%%%
As already stressed, the orbit space of a CLG can be characterized through
an associated \wPm\, and the specification of the surface ${\cal Z}$,
in case of non-coregularity.
%%%%
Our final aim is the determination and classification of the orbit
spaces of the CLG's, avoiding to pass through an explicit determination
of a MIB for each group.

For that reason,
following the same approach proposed in \cite{681}, and developed in
\cite{682}, for coregular groups,
we shall now look at the boundary conditions from a different point of
view.
We shall forget altogether the group $G$ and the space $\real^{n}$,
and we shall think of $\{p_1,\dots ,p_q\}$  as a set of {\em weighted
indeterminates}, with integer weights $d_1, \dots ,d_q$ satisfying the
following conditions:

\eqll{d_{1} \geq d_{2} \geq \ldots d_{q} \,=\, 2\, .}{a3a}
%%%%%%%%%%%%%%%%%%%%%%%%%%%%%%%%%
We shall associate {\em formal}
\wPms\ satisfying condition {\bf P1} in \S\ref{P12}
to a weighted set $\{p\}=\{p_1,\dots ,p_q\}$ .

\begin{definition}   \label{Pmatrix}
Let $\{p\}=\{p_1,\dots ,p_q\}$ be a set of weighted real variables of
weights
$\{d\}=\{d_1,\dots ,d_q\}$ $(d_1\ge\dots\ge d_q=2)$. We shall say that
$\wP(p)$ is a {\em (formal) $\wP$-matrix }
associated with $\{p\}$, if it satisfies the following conditions:

\begin{itemize}
\item[i)] $\wP(p)$ is a real, symmetric, $q\times q$ matrix.

\item[ii)] The matrix elements $\wP_{ab}(p)$ are $w$-homogeneous
polynomials in $p$ and their weights are
$w(\wP_{ab})=d_a+d_b-2,\ a,b=1,\dots ,q.$

\item[iii)] $P_{qa}(p)=2d_ap_a,\ a=1,\dots ,q.$

\end{itemize}
\end{definition}

The equivalence of two formal \wPms\ will be defined as in \S~\ref{Cosclg}.
%%%%%%%%%%%%%%%%%%%%%%%%%%%

The boundary relations will be considered as  {\em equations} in which
the polynomial functions involved play the role of unknown polynomial
functions of $p$.
With the above meaning for the symbols, Eq.~(\ref{mas}) will be called
the {\it master equation} and its canonical form (Eq.~(\ref{can})) {\em
the canonical equation}.

We shall be interested in solutions $\left( \wP(p),A(p)\right)$ of the
canonical equation satisfying some additional conditions, which are certainly
satisfied by all \wPms\ originating from CLG's and associated to convenient
MIB's.

\begin{definition}   \label{Ppermessa}
A \wPm\ $\wP(p)$ associated to the weighted set of variables
$\{p\}=\{p_1,\dots ,p_q\}$  of weights
$\{d\}=\{d_1,\dots ,d_q\}$ $(d_1\ge\dots\ge d_q=2)$,
will be said to be  {\em allowable of
r-type $(q,k)$ }
if it satisfies the following conditions:

\begin{itemize}

\item[i)] There is a $k$-dimensional $(k \geq 1)$ irreducible algebraic
surface ${\cal Z}$ in $\real^q$ whose defining equations $\widehat{F}_A(p) =
0$ can be expressed in terms of irreducible $w$-homogeneous polynomials
$\widehat{F}_A(p)$, such that $\partial \widehat{F}_A(0) = 0$.  On $\cal Z\;$,
$\mbox{{\rm rank}}\, (\wP(p))\le k$ and the set ${\cal R} = \{p \in {\cal Z}
\,|\, \wP(p) \geq 0\,,\, \mbox{{\rm rank}}(\wP(p)) = k \}$ is
$k$-dimensional and connected; the closure $\overline {\cal R}$ of $\cal R$
coincides with the set ${\cal R}^{(\geq)} = \{p \in {\cal Z} \,|\, \wP(p)
\geq 0 \}$.

\item[ii)] $\wP(p)$ satisfies the boundary conditions  {\rm (\ref{A2})},
for each primary stratum ${\sigma_{i}}^{(\alpha)}$ of
$\overline{\cal R}$, and the second order boundary condition
{\rm (\ref{FE1})} for each $\widehat{F} (p) \in {\cal I}({\cal Z})$.
\end{itemize}
\end{definition}

In the following, the expression {\em allowable \wPm\ } will be abbreviated
in $A \widehat P M$.

\begin{remark}
{\rm
The conditions under item {\em i)} in the Def.~(\ref{Ppermessa}) imply
immediately that for an $A \wP M$, $\,\mbox{{\rm rank}}\,(\wP(p))< k$ on the
boundary $\,\overline{{\cal R}} \setminus {\cal R}\,$ of ${\cal R}$.  }
\end{remark}

\begin{remark} {\rm In the first definition of $A\wP M$'s
\cite{682}, a condition of compactness of the set $\Pi \cap \overline{{\cal
R}}$ was included.  From the proof of Theorem 7.1 in \cite{682}, it is easy to
realize that the set $\Pi \cap \overline{{\cal R}}$ is compact for every
formal $\wP$-matrix, as a consequence of its structure.  } \end{remark}

The $A \widehat P M$'s of r-type $(q,q)$, that could be associated to
coregular groups, have all been determined in \cite{681,682,685}
for $q \leq 4$, solving the canonical equation with the initial conditions
specified in
(\ref{a2b}) and (\ref{iniP}). We shall call these solutions (and the
associated \wPms ), {\em proper solutions (proper \wPms) of r-type} $(q,q)$.

Analogously, in the following
section we shall determine the $A \widehat P M$'s
that could be associated to non-coregular groups of r-type $(q,q-1)$
from the solutions of
the canonical equation which are proper of r-type $(q,q-1)$ in the sense
specified by the following definition:

\begin{definition}
Let $\wP(p)$ be a \wPm\ associated to the weighted set of variables
$\{p_1,\ldots,p_q\}$ and
$\left( \wP(p), \widehat{F}(p)\right )$ be a solution of the canonical
equation. The couple $\left( \wP(p), \widehat{F}(p)\right )$
will be said to be a {\em proper solution
of r-type $(q,q-1)$} if it satisfies the following conditions:
\begin{itemize}

\item[i)] $\widehat F(p)$ is irreducible on the complex field,

\item[ii)] $\partial \widehat{F}(0) = 0$,

\item[iii)] There are $w$-homogeneous polynomials $L_a(p)$ such that
Eqs.~{\rm (\ref{S1})} and {\rm (\ref{S2})} are satisfied.

\end{itemize}
A \wPm\ obtained from a proper solution will be said to be {\em
a proper \wPm\ (henceforth abbreviated in $P \widehat P M$)}.
\end{definition}
In the following section for each proper solution of the canonical equation we
shall determine the conditions under which the associated \wPm\ is allowable.
In order to determine the semi-positivity domain of $\wP(p)$, we shall make
use of the following well known theorem and of a lemma which we shall prove
below;

\begin{theorem}\label{T2}
Let $A$ be a real symmetric matrix. Then $A\,\ge 0$ if and only if
$A_{\alpha} \geq 0$ for all $\alpha$, where $\left \{ A_{\alpha}
\right \}$ is the set of determinants of principal (i.e., symmetric)
minors of $A$.
\end{theorem}

%Taking into account the $w$-homogeneity properties of the elements of
%$\wP(p)$, for further reference, it is easy to realize that

\begin{lemma}  \label{Remark}
Let $(\wP(p), \widehat{F}(p))$ be a proper solution of r-type $(q,q-1)$
and ${\cal Z} = \{p \in \real^{q} \,\mid\, \widehat{F}(p)=0 \}$, then,
with the same meaning of the symbols as in {\rm Def.~(\ref{Ppermessa})},
the following two conditions are necessary and sufficient for
$\wP(p)$ being an $A \widehat P M$ of r-type $(q,q-1)$:
\begin{itemize}
\item[i)] The semi-algebraic set ${\cal R}_{1}= \Pi \cap
{\cal R}$ is $(q-2)$-dimensional and connected and its closure $\overline{
{\cal R}_{1}}$ coincides with the semi-algebraic set ${\cal R}_{1}^{\ge} = \Pi
\cap {\cal R}^{\ge}$.
\item[ii)] $\wP(p)$ satisfies the boundary conditions  {\rm (\ref{A2})}
at every singular primary stratum of ${\cal R}^{(\geq)}$.
\end{itemize}
\end{lemma}
%%%%%%%%%%%%%%%%%%%%  Dimostrazione  %%%%%%%%%%%%%%%%%%%%%%%%%%%%%%%%
\noindent{\em Proof:}
%\begin{proof}
Since in the statement of the lemma item {\em ii)} essentially coincides
with item {\em ii)} in Def.~(\ref{Ppermessa}), it will be sufficient to note
the following facts.
\begin{itemize}
\item[a)] From $\wP_{qq}(p) = 4 p_q$ and Theorem \ref{T2} it follows that
$\wP(p) \geq 0$ only for $p_q \geq 0$.

\item[b)] Owing to $w$-homogeneity properties of
$\widehat{F}(p)$, the point $p=0$ belongs to ${\cal Z}$
and $\wP(0)=0$. The origin of $\real^{q}$ is the only point where
$\mbox{{\rm rank}}\,(\wP(p)) = 0$;
moreover, if $p \in {\cal Z}$, then
$\left(  s^{d_1}p_1,\ldots,s^{d_{q-1}}p_{q-1},s^2 p_q \right) \in
{\cal Z}$ for all $s \in \real$.

\item[c)] Let us denote by $r:\real^{q} \longrightarrow \Pi$,
the map such that $p \equiv (p_1,\ldots,p_{q-1},p_q) \mapsto
r(p) \equiv
\left( {\displaystyle \frac{p_1}{{p_q}^{d_1/2}},\frac{p_2}{{p_q}^{d_2/2}},
\ldots,\frac{p_{q-1}}{{p_q}^{d_{q-1}/2}},1} \right)$,
which is well defined $\forall p_{q} > 0 $.

Then, the $w$-homogeneity properties of $\wP(p)$ assure that for $p_q >0$:
\eqll{p_q \wP(p) \;=\;  T \, \wP(r(p)) \,T \,,}{trapa}
where $T=\mbox{\rm{diag}}({p_q}^{d_1/2},{p_q}^{d_2/2},
\ldots,{p_q}^{d_{q-1}/2},p_q)$.
As a consequence, $\wP(p) \geq 0$ if and only if $\wP(r(p)) \geq 0$,
and, using also the $w$-homogeneity properties of $\widehat{F}(p)$
mentioned under item {\em b),} the following relations hold true:
\eqll{ {\cal R} \,=\, \left \{
\left( s^{d_1} r_{1}(p),\ldots,s^{d_{q-1}} r_{q-1}(p), s^2 \right) \;\mid\;
r(p) \in {\cal R}_{1} \,,\,s>0 \right \} \,,}{trapa1}
%%%%
\eqll{ {\cal R}^{(\geq)} \,=\, \left \{
\left( s^{d_1} r_{1}(p),\ldots,s^{d_{q-1}} r_{q-1}(p), s^2 \right) \;\mid\;
r(p) \in {\cal R}_{1}^{(\geq)} \,,\,s\geq 0 \right \} \,.}{trapa2}
\end{itemize}

According to the above remarks, it is now easy to realize that ${\cal R}$ and
${\cal R}^{(\geq)}$ are respectively homeomorphic to ${\cal R}_{1} \times
\real_{+} $, and ${\cal R}^{(\geq)}_1 \times\overline{\real_{+}}$, so that
we can conclude that:

\begin{enumerate}
\item $\cal R$ is $(q-1)$-dimensional and connected iff ${\cal
R}_1$ is $(q-2)$-dimensional and connected;
\item $\overline{\cal R}\ =\ {\cal R}^{\ge}\;$ iff $\;\overline{\cal R}_1\ =\
{\cal R}_1^{\ge}$. \scat
\end{enumerate}
%\end{proof}
\vskip3truemm

%%%%%%%%%%%%%%%%%%%%%%%%%%%%%%%%%%%%%%%%%%%%%%%%%%%%%%%%%%%%%%%%%%%%%%%%%

The precise correspondence between $P \widehat P M$ and
$A \widehat P M$ and between formal $A \widehat P M$
and \wPms\ originating from CLG's has not yet been fully
clarified. The following facts have however been proved in \cite{682,685}:

\begin{itemize}
\item[i)] the $P \widehat P M$'s of r-type $(q,q)$ are necessarily
$A \widehat P M$'s of r-type $(q,q)$;

\item[ii)] For $q\leq 4,$ $\,P \widehat P M$'s of r-type $(q,q)$
have been shown to be $\geq 0 $ only on a connected
$q$-dimensional semi-algebraic subset of ${\cal Z}=\real^{q}$
and, for most of them, the boundary conditions have been checked.
If this result could
be shown to hold in general, there would be identity between
$A \widehat P M$'s and $P \widehat P M$'s of r-type $(q,q)$.

\item[iii)] For each choice of the degrees $\{ d_{1}, d_{2}, \ldots, d_{q} \}$,
there exists only a finite (or null) number of non equivalent \wPms\
of r-type $(q,q)$, at least for $q \leq 4$. This implies that the possible sets
of degrees are limited by {\em selection rules}.
 All of these $P \widehat{P} M$'s can be organized in {\em towers} and the
degrees of the elements of the same tower can be written in the form
$d_\alpha=sd_\alpha^ {(0)},\ \alpha= 1,\dots ,q-1,$ where $s$ is a positive
integer scale parameter. All the \wPms\ of the same tower coincide for
$p_q=1$.

\item[iv)] Any \wPm\ originating from a coregular CLG with no fixed points is
necessarily equivalent to an $A \widehat{P} M$ of r-type $(q,q)$.
As a consequence, the selection rules on
the sets of allowable degrees mentioned under item iii) hold true for all
coregular CLG's.  At present we do not know whether the converse holds also true,
i.e., if every $A \widehat{P} M$ of r-type $(q,q)$ is generated by a
coregular CLG with no fixed points.  A partial answer to this question has
however been given in \cite{687} and \cite{688}, where it has been checked that
the \wPms\ originating from all the finite coregular groups (which are
the groups generated by reflections \cite{Chev,Shep}) and from all the coregular
 representations of compact simple Lie groups,
with less than $5$ basic invariants, can be found among the
$A \widehat{P} M$'s
listed in \cite{682,685}.
\end{itemize}

\vskip3truemm
The correspondence between $P \widehat P M$'s and
$A \widehat P M$'s of r-type $(q,q-1)$ is much more difficult
to study, for general values of $q$.  Here we shall limit ourselves to note
that, if $\wP (p)$ is a \wPm\ of a CLG  of r-type $(q,q-1)$ and
$\widehat F (p) = 0$ is the (basic) relation among the elements
of the MIB $\{p\}$,
then, in $F$-bases $\{p' \}$,
the couple $(\wP'(p'),\widehat{F}'(p'))$ is
necessarily a {\em proper} solution of r-type $(q,q-1)$ of the
canonical equation.

In the following section we shall start this analysis by determining all
the proper \wPms\ of r-type $(q,q-1)$, in the simplest case $q=3$.

%%%%%%%%%%%%%%%%%%%%%%%%%%%%%%%%%%%%%%%%%%%%%%%%%%%%%%%%%%%%%%%%%%%%%%

\section{Non-coregular groups of r-type $(3,2)$}
In this section we shall determine the \wPms\ (and therefore the isotropy
classes of the orbit spaces) of all the non-coregular groups of r-type $(3,2)$,
i.e., with $3$ basic polynomial invariants connected by only one independent
relation. We shall start by determining the $P \widehat{P}M$'s of r-type
$(3,2)$; from these we shall select the allowable ones.

\subsection{Proper and allowable \wPms\ of r-type $(3,2)$ }
We shall limit ourselves to sketch the procedure we have followed to determine
all the $P\widehat{P}M$'s of r-type $(3,2)$.  The most general form allowed by
the conditions listed in \S\ref{P12} under {\bf P1} and {\bf P2} for the
elements of the matrix $\wP (p)$ and by the weights of the $L_a(p)$ is the
following:

\eqll{\begin{array}{rcl}
\wP_{11}(p)&=& d_1^2[p_1 a_1(p_2,p_3) +a_2(p_2,p_3)] \\
\wP_{12}(p)&=& d_1d_2[p_1 b_1(p_3) + a_3(p_2, p_3)] \\
\wP_{22}(p)&=& d_2^2[p_1 b_2(p_3) + p_2 b_3(p_3) + b_4(p_3)]\\
\wP_{a3}(p)&=& 2d_ap_a\\
L_1(p)&=& d_1a_4(p_2, p_3)\\
L_2(p)&=& d_2b_5(p_3)\,.
\end{array}
}{ini}
where the $a$'s and $b$'s are unknown polynomial functions
whose weights, determined according to (\ref{weightP}) and (\ref{weightL}) are
specified in Table \ref{table1}. The factorization of the $d_i$'s  is
suggested by the structure of the equations we shall have to solve.

\begin{table}[h]
\begin{center}
\begin{tabular}{||cc||cc||cc||} \hline
polynomial&  weight   &polynomial & weight     & polynomial & weight  \\\hline
$a_1$    & $d_1 - 2$ & $a_2$     & $2d_1 - 2$ &$a_3$     & $d_1 + d_2 - 2$ \\
$a_4$    & $d_1 - 2$ &$b_1$  & $d_2 - 2$  & $b_2$    & $2 d_2 - d_1 - 2$\\
$b_3$    & $d_2 - 2$ &$b_4$  & $2 d_2 - 2$ & $b_5$     & $d_2 - 2$ \\
\hline
\end{tabular}
\end{center}
\caption{Weights of the unknown polynomials entering in the definition of $\wP
(p)$}
\label{table1}
\end{table}

Since $\widehat F(p)$ has been required to be an irreducible (on the complex numbers)
polynomial in
the indeterminates $p_1,\dots ,p_q$, its gradient $\partial\widehat F(p)$
cannot vanish identically on the surface
\eqll{ {\cal Z} = \left \{ p \in \real^{q} \; \mid\; \widehat{F}(p) \,=\,0
\right \} \,.}{supzeta}
Therefore, thinking of the canonical equation as a system of linear
equations, it is easy to realize that the
determinant of the matrix $\wP(p)$ of the coefficients has to vanish where
$\widehat F(p)$ vanishes.
This means that $\widehat F(p)$ is necessarily a factor of det$\,\wP(p)$ and,
consequently:

\eqll{w(F) \le w({\rm det}\,\wP) = \summ 13a (2d_a - 2).}{w(F)}

The most general form allowed for $\widehat F(p)$ is therefore the
following:

\eqll{\widehat F(p) =  f_3(p_3)\,p_1^3 + f_2(p_2,p_3)\,p_1^2 +
f_1(p_2,p_3)\,p_1 + f_0(p_2,p_3),}{F}
where the $f$'s are $w$-homogeneous polynomials of weights

\eqll{w(f_j) \, =\, w(F) - jd_1,\qquad j=0,\dots ,3}{wf}
and the condition $\partial \widehat{F}(0) = 0$ requires:
\eqll{f_1(0,0)\,=\, 0,\qquad \partial f_0(0,0)\,=\, 0\,.}{dF}
The overall normalization of $\widehat{F}(p)$ can be fixed arbitrarily.

The couple $(\wP(p),\widehat F(p))$ has to satisfy the canonical equation
(\ref{can}) and the additional conditions (\ref{S1}), (\ref{Lq}) and
(\ref{S2}).  The derivative $\partial_3 \widehat{F}(p)$ can be eliminated from
these equations making use of the $w-$homogeneity condition on
$\widehat{F}(p)$:

\eqll{\summ 1{2}\alpha d_\alpha  p_\alpha  \partial_\alpha\widehat F(p) +
2\, p_3 \partial_3 \widehat F(p) = w(F)\,\widehat F(p)\,.}{hom}

In this way one obtains for the canonical equation (\ref{can}):
\eqll{\summ 12{\beta }\left(p_3\wP_{\alpha\beta}(p) - d_\alpha
d_\beta p_\alpha p_\beta \right)\partial_\beta \widehat F(p)
+ w(F) d_\alpha p_\alpha \widehat F(p)\, =\, 0\,,\qquad \alpha =1,2
}{can3}
and for the additional condition (\ref{S1}):
\eqll{\summ 12\alpha \left(p_3 L_\alpha (p) - D d_\alpha p_\alpha\right)
\partial_\alpha \widehat F(p) + Dw(F)\widehat F(p) \, =\, 0\,.
}{S3}

The solution of (\ref{can3}) and (\ref{S3}),
fulfilling the condition $\partial \widehat{F}(0)=0$ ,
can be obtained through the following
steps:

\begin{enumerate}

\item Since in (\ref{can3}) and (\ref{S3})
there are no derivatives with respect to $p_3$,
and the equations are $w$-homogeneous, it is advantageous to solve them
first for $p_3 = 1$ and to reintroduce the dependence on $p_3$
in the solutions at the end. Then, let us set:

\eqll{f_i(p_2,1)\ =\ f_i(p_2),\ i=0,1,2;\qquad f_3(1)\ =\ f_3;}{effe}

\eqll{a_i(p_2,1)\ =\ a_i(p_2),\ i=1,\dots ,4;\qquad b_j(1)\ =\ b_j,\ j=1,
\dots ,5}{ab}

\item The dependence on $p_1$ is made explicit after substituting in
(\ref{can3}) and (\ref{S3}) the expressions (\ref{ini}) and (\ref{F}) for
$\wP_{ab}(p)$ and $\widehat F(p)$. Therefore,
the principle of identity for polynomials allows to eliminate easily
the variable $p_1$. One obtains in this way the following system of coupled
algebro-differential equations, where $w$ stands for $w(F)$:

\noindent $
%ee[1,4] =
(w - 3 d_1) f_3\ =\ 0\,,$ \\

\noindent $
%ee[1,3] =
3 d_1 f_3 a_1(p_2) + (w - 2 d_1) f_2(p_2) +
  d_2 (b_1 - p_2) f_2'(p_2)\ =\ 0\,,$  \\

\noindent $
%ee[1,2] =
3 d_1 f_3 a_2(p_2) +
  2 d_1 a_1(p_2) f_2(p_2) + (w - d_1) f_1(p_2) + d_2 a_3(p_2) f_2'(p_2)
%%%rep%%% \\ \qquad\qquad\qquad\qquad\qquad\qquad\qquad\qquad\qquad\qquad
+d_2 (b_1 - p_2) f_1'(p_2)\ =\ 0\,,$  \\

\noindent $
%ee[1,1] =
2 d_1 a_2(p_2) f_2(p_2) +
  d_1 a_1(p_2) f_1(p_2) + w f_0(p_2) +
  d_2 a_3(p_2) f_1'(p_2) +
  d_2 (b_1 - p_2) f_0'(p_2)\ =\ 0\,,$  \\

\noindent $
%ee[1,0] =
d_1 a_2(p_2) f_1(p_2) +
  d_2 a_3(p_2) f_0'(p_2)\ =\ 0\,,$  \\

\noindent $
%ee[2,3] =
(3 b_1 d_1 + w p_2 - 3 d_1 p_2) f_3 +
  b_2 d_2 f_2'(p_2)\ =\ 0\,,$  \\

\noindent $
%ee[2,2] =
3 d_1 f_3 a_3(p_2) + (2 b_1 d_1 +
  w p_2 - 2 d_1 p_2) f_2(p_2) +
  d_2 (b_4 + b_3 p_2 - p_2^2)
f_2'(p_2) +
  b_2 d_2 f_1'(p_2)\ =\ 0\,,$  \\

\noindent $
%ee[2,1] =
2 d_1 a_3(p_2) f_2(p_2) +
  (b_1 d_1 + w p_2 - d_1 p_2) f_1(p_2) +
  d_2 (b_4 + b_3 p_2 - p_2^2)
f_1'(p_2)
%%rep%%%  \\ \qquad\qquad\qquad\qquad\qquad\qquad\qquad\qquad\qquad\qquad\qquad\qquad
+ b_2 d_2 f_0'(p_2)\ =\ 0\,,$  \\

\noindent $
%ee[2,0] =
d_1 a_3(p_2) f_1(p_2) + w p_2 f_0(p_2) +
  d_2 (b_4 + b_3 p_2 - p_2^2)
f_0'(p_2)\ =\ 0\,,$ \\

\noindent $
%Su[3] =
D (w - 3 d_1) f_3\ =\ 0\,,$  \\

\noindent $
%Su[2] =
3 d_1 f_3 a_4(p_2) + D (w - 2 d_1) f_2(p_2) +
  d_2 (b_5 - D p_2) f_2'(p_2)\ =\ 0\,,$  \\

\noindent $
%Su[1] =
 d_1 a_4(p_2) f_2(p_2) + D (w - d_1) f_1(p_2) +
  d_2 (b_5 - D p_2) f_1'(p_2)\ =\ 0\,,$ \\

\noindent $
%Su[0] =
d_1 a_4(p_2) f_1(p_2) + D w f_0(p_2) +
  d_2 (b_5 - D p_2) f_0'(p_2) \ =\ 0\,. $\\

\item The solution of the system of equations just written
is much more lengthy and trickier. In principle, it could
be reduced to the solution of a system of algebraic equations by expanding
the unknown polynomial functions in powers of $p_2$ and identifying to $0$
the coefficients of homonymous powers of $p_2$ in each equation. It has to
be recalled however, that the weights of the polynomials are functions of the
degrees $d_i$, which are parameters. So, the high number of
variables one is obliged to introduce in this way and the high number of
coupled algebraic equations to solve make this standard procedure quite
difficult to handle.

The easiest way to obtain the solutions seems to be through a combined use
of algebraic and integro--differential methods, with a clever choice of the
order in which to solve the various equations.
\end{enumerate}

It would be too long to describe the details of the calculations that led to
the determination of all the solutions of the system of equations
(\ref{can3}), (\ref{S3}) and (\ref{dF}). We shall
limit ourselves to report the solutions, discarding those in which $\widehat
F(p)$ turns out to be reducible on the complexes. The proper solutions will be
collected in 3 families, S1, S2 and S3, corresponding respectively to the
degrees
$(d_1,d_2)$ = $(k(2m + 1),2k)$,
%$m,k\in\natural$,
$(d_1,d_2)$ = $(6k,4k)$,
%$k\in\natural$
and $(d_1,d_2)$ = $(k+1,k+1)$,
%with $\natural \ni k\ge 2$.
where $m$ and $k$ are positive integers. Each family will be discussed
separately in each of the following three subsections. For each family we
shall determine the number of distinct equivalence classes of \wPms\, picking
up for each class a representative \wPm, chosen so that the
numerical coefficients of all the polynomial involved are integer numbers. The
allowability conditions will finally be checked for each representative \wPm.

For all the solutions, the overall normalization of $\widehat
F(p)$ will be chosen so that the coefficient of the highest power of $p_1$,
which turns out to be always a constant, equals 1.

\subsubsection{Solution S1}
The family S1 of proper solutions is found in correspondence with the
degrees

\eqll{ d_1=k(1+2m)\,, \qquad d_2=2k\,, \qquad k,m \in
\natural_{*}\,.}{degrees1}

In the rest of this section, $k$ and $m$ will be considered as fixed.
For the unknown polynomial functions $L_1, L_2,\widehat F,\wP_{ij}$, $i,j=1,2$
one finds the following expressions, in terms of two real parameters,
$b_3\ne 0$ (defined in (\ref{ini})) and $c$
(originating as an integration constant):

\eqll{ L_1\ =\ 0;\qquad L_2(p_3)\ =\ d_2 D b_3 \,p_{3}^{k-1}\,.}{elle1}

\eqll{ \widehat{F}=p_{1}^{2} + c(p_{2} - b_{3} p_{3}^k)^{2m + 1}, }{Fsol1}

\eqll{\begin{array}{rcl}
\wP_{11}(p) &=&
d_{1}^{2}\, {b_3}\, c\, p_3^{k-1} \left( p_2 - b_3
p_3^k \right)^{ 2 m} \\
&&\\
\wP_{12}(p) &=&
  0\\
&&\\
\wP_{22}(p) &=&
   d_2^{2}\, b_3 \, p_2\, p_3^{k-1}\\
   \end{array}
}{matPsol1}

Different values of the parameters $b_3$ and $c$ do not necessarily determine
non-equivalent \wPms. In fact,

\begin{proposition}
For each fixed choice of $(m,k)$, the \wPms\ determined by the family S1
of proper solutions form two distinct classes of equivalent \wPms.
\end{proposition}

\noindent{\em Proof:}
By means of the following formal MIBT

\eqll{\begin{array}{rcl}
p'_1 &=&  \left | b_3^{2m+1} c \right |^{- \frac{1}{2}} p_1 \\
&&\\
p'_2 &=& b_3^{-1} \, p_{2}  \\
&&\\
p'_3 &=& p_3\,;
\end{array}
}{MIBTt2}
after setting
$$\epsilon = \mbox{sign} (b_3 c),$$
one obtains from (\ref{equiv}), (\ref{Fsol1}), (\ref{matPsol1}) and
(\ref{degrees1}):

$$\widehat F'(p') \,=\, {p'_1}^2 + \,\epsilon  \left(
            {p'_2} - {p'_3}^{k} \right)^{2m+1} \;,$$

\eqll{\begin{array}{rcl}
\wP'_{11}(p') &=&
d_{1}^{2}\; \epsilon \,{p'_3}^{k - 1}
\left( p'_2 - {p'_3}^{k} \right)^{2m}\,  \\
&&\\
\wP'_{12}(p') &=&
0  \\
&&\\
\wP'_{22}(p') &=&
   d_{2}^{2}\; {p'_2}\, {p'_3}^{k - 1} ,
\end{array}
}{matP'sol1}
and

$${\rm det}(\wP'(p'))\,=\, -4\, d_{1}^{2}\, d_{2}^{2}\; {p'_2}{p'_3}^{k - 1}
 \left \{ {p'_1}^{2} - \epsilon(        {p'_3}^{k} - {p'_2} )^{2m+1}
  \right\}
.$$
In this way all the arbitrary parameters, but for $\mbox{sign} (b_3 c)$,
have been absorbed in the redefinition of the MIB.
It is now evident that no MIBT can cause, as unique consequence, a change of
sign in front of $\epsilon$ in (\ref{matP'sol1}).

For each fixed choice of $(k,m)$, there are therefore two
distinct classes of  equivalent \wPms, represented by the \wPms\
defined in (\ref{matP'sol1}), respectively for $\epsilon =1$ and $\epsilon
=-1$.    \scat

\begin{proposition}
The proper solutions of the family S1 do not determine allowable \wPms.
\end{proposition}

\noindent{\em Proof:}
It will be sufficient to check the allowability
conditions on the representative \wPms\ defined in (\ref{matP'sol1}).
To this end, let us determine the semi-positivity domain of $\wP'(p')$ in
${\cal Z}$, separately for $\epsilon =\pm 1$.

For $\epsilon=-1$ and $p'_{3}=1$, from  Theorem \ref{T2} and Lemma
\ref{Remark} we  immediately realize that $\wP'(p')\ge 0$ only at $p'_2=1$,
$p'_1=0$, which is a $0$-dimensional set.
Therefore, for $\epsilon=-1$, the matrix $\wP'(p')$ is not allowable.

For $\epsilon=+1$, the subset ${\cal R}_1^{(\ge)}\subset\Pi \,\cap \,{\cal
Z}\;$ where $\wP'(p') \geq 0 $ is determined by the following conditions:
\eqll{
{\cal R}_{1}^{(\geq)}=\Pi \cap {\cal R}^{(\geq)} =
 \left \{ (p'_1,p'_2) \; \mid \;
{p'_1}^{2} = (1 - {p'_2})^{2m+1}, \hspace{2em} 0\leq {p'_2} \leq 1
\right \} , }{xyzh}
where, obviously, $\mbox{{\rm rank}}\,(\wP'(p'))=2$ in the interior of
$ {\cal R}_{1}^{(\geq)}$ and $\mbox{{\rm rank}}\,(\wP'(p'))=1$ on the
boundary. The variety ${\cal Z}$ admits the following
parametric representation:
\begin{eqnarray}
p'_1 &=& t^{2m+1} \nonumber \\
&& \\
p'_2 &=& 1-t^2\,\nonumber
\end{eqnarray}
and ${\cal R}_{1}^{(\geq)}$ corresponds to values of $t \in [-1,1]$, the
singular strata corresponding to $t=0,\,\pm 1$.

The set of points ${\cal R}_{1}$ where $\wP'(p') \geq 0 $ and $\mbox{{\rm
rank}}\, (\wP(p)) =2$ is not connected, as it is evident from Fig.~(1).  We
conclude therefore that, owing to Lemma \ref{Remark}, the two distinct classes
of equivalent \wPms\, obtained from the family S1 of proper solutions are
{\em not allowable.}  \scat

We have reported in Table \ref{table3} the equations defining the singular
primary strata $\sigma^{(i)}$ of ${\cal R}^{(\geq)}$ and a possible choice for
the generators of the associated ideals ${\cal I}(\sigma^{(i)})$.

It is also easy to check that the boundary conditions are not satisfied at the
singular primary strata $\sigma^{(2)}$ and $\sigma^{(3)}$ defined in Table
\ref{table3}.

%%%%%%%%%%%%%%% Tabella 2  %%%%%%%%%%%%%%%%%%%%%%%
%%    (table 3) il riferimento %%%%
%%%%%%%%%%%%%%%%%%%%%%%%%%%%%%%%%%%%%%%%%%%

\begin{table}
\begin{center}
\begin{tabular}{||c|c|c|c|c||} \hline \hline
stratum    &  $t$  &defining relations & generators & boundary conditions
\\  \hline
  &  & $p'_1\,=0$ & $f^{(1)}\,=\,{p'_1}$  &    \\
$\sigma^{(1)}$ &  $0$  &   &    & satisfied \\
   &   &  $ p'_2={p'_3}^{k}$ &  $f^{(2)}\,=\,p'_2 - {p'_3}^{k}$ &  \\ \hline
   &    & ${p'_1}^{2}\,={p'_{3}}^{k (2m+1)} $ &
     $f^{(1)}\,=\,{p'_1}^{2}-{p'_{3}}^{k (2m+1)}$ &
     \\
$\sigma^{(2)}$ &  $1$ & $p'_2\,=\,0$       &     & not satisfied  \\
  &  & $p'_1\,>\,0$ & $f^{(2)}\,=\,p'_2 $ &
 \\  \hline
   &    & ${p'_1}^{2}\,={p'_{3}}^{k (2m+1)} $ &
     $f^{(1)}\,=\,{p'_1}^{2}-{p'_{3}}^{k (2m+1)}$ &
     \\
$\sigma^{(3)}$ &  $-1$ & $ p'_2\,=\,0 $     &     & not satisfied  \\
  &  & $p'_1\,<\,0$ & $f^{(2)}\,=\,p'_2 $ &
 \\ \hline \hline
\end{tabular}

\end{center}
\caption{Defining relations for the singular primary strata, generators and
boundary conditions of solution S1 in the case $\epsilon \,=\,+1$.}
\label{table3}
\end{table}

%%%%%%%%%%%%%%%%%%%%%%%%%%%%%%%%% Figura 1 %%%%%%%%%%%%%%%%%%%%%%%%%

\begin{figure}
\centerline{ \epsfxsize=3.2cm \epsffile{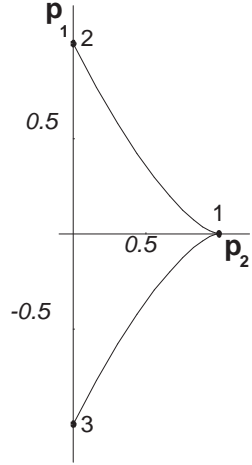} }
\caption[]{Stratification of ${\cal R}_{1}^{(\geq)}$ for solution
S1. The numbers $j$, $j=1,2,3\;$, label the singular primary
strata $\sigma^{(j)}$ defined in Table 2.}
\end{figure}

%%%%%%%%%%%%%%%%%%%%%%%%%%%%%%%%%%%%%%%%%%%%%%%%%%%%%%%%%%%%%%%%%%%%

\subsubsection{Solution S2}
The solution S2 is found in correspondence with the degrees

$$d_1 = 6 k, \qquad d_2 = 4 k, \qquad k\in\natural_*$$
and is defined by the following expressions of the unknown polynomial functions:

\eqll{L_1(p_2,p_3) \,=\, {d_1 D b_1 \over 4 b_2}\,
\left [ 2 p_2+(3 b_1 +2 b_3)p_{3}^{2k} \right ]\,p_{3}^{k-1}\,;
\qquad L_2(p_3) \,=\,
{d_2 D \over 2 }(3 b_1 + 2 b_3) p_{3}^{2k-1} \,;}{sxh2}

\eqll{\begin{array}{rcl}
\widehat F(p) &=&  \displaystyle{\frac{1}{8 b_2^2}}
\left\{8 b_2^2 p_1^2
+ 4 b_1 b_2 p_1 p_3^k [(3 b_1 + 2 b_3) p_3^{2 k}- 6 p_2)]
+ b_1 [ 4 p_2^3 - 12 b_3 p_2^2 p_3^{2 k}+
\right. \nonumber\\
&&\\
&& \left.  (9 b_1^2 + 24 b_1 b_3 + 12 b_3^2) p_2 p_3^{4 k}
- (9 b_1^3 + 21 b_1^2 b_3 + 16 b_1 b_3^2 + 4 b_3^3) p_3^{6 k}
\right\},
\end{array}
}{Fsol2}

\eqll{\begin{array}{rcl}
\wP_{11}(p)  &=&
{\displaystyle b_1 d_1^2\over\displaystyle  16 b_2^2}\, p_3^{k - 1} \left\{
4 b_2 p_1 \left[ 2 p_2 + (b_1 - 2 b_3) p_3^{2 k} \right]
+ 4 (b_1 + 2 b_3) p_2^2 p_3^{k} - \right. \nonumber\\
&& \left. 8 b_3 (b_1 + 2 b_3) p_2 p_3^{3 k}
+ (3 b_1^3 + 14 b_1^2 b_3 + 20 b_1 b_3^2 + 8 b_3^3) p_3^{5 k}
\right\}  \\
&&\\
\wP_{12}(p)  &=&
{\displaystyle b_1 d_1 d_2\over\displaystyle 8 b_2}\, p_3^{2 k - 1} \left\{ 8 b_2 p_1
+ 6 (b_1 + 2 b_3) p_2 p_3^k
- (3 b_1^2 + 8 b_1 b_3 + 4 b_3^2) p_3^{3 k}
\right\}   \\
&&\\
\wP_{22}(p)  &=&
{\displaystyle d_2^2\over\displaystyle 4} \, p_3^{k - 1} \left\{
4 b_2 p_1
+ 4 b_3 p_2 p_3^k
+ (3 b_1^2 + 2 b_1 b_3) p_3^{3 k}\right\} ;
\end{array}
}{matPsol2}
the determinant of $\wP(p)$ is

\eqll{\begin{array}{rcl}
\mbox{{\rm det}}(\wP (p)) &=&
{\displaystyle \frac{d_1^2 d_2^2}{4 b_2^2}} \, p_3^{k - 1}
   \left(2 b_2 p_1 + b_1 p_2 p_3^k + 2 b_3 p_2 p_3^k\right)
\left\{ -8 b_2^2 p_1^2 +
4 b_1 b_2 \left[6 p_2 - \right.\right. \nonumber\\
&&\\
&& \left.\left. (3 b_1 + 2 b_3) p_3^{2 k}\right] p_1 p_3^k
+ b_1 \right(- 4 p_2^3 + 12 b_3 p_2^2 p_3^{2 k}
- 3 (3 b_1^2 + 8 b_1 b_3 + \nonumber\\
&&\\
&&\left.\left.
4 b_3^2) p_2 p_3^{4 k}
+ (9 b_1^3 + 21 b_1^2 b_3 + 16 b_1 b_3^2 + 4 b_3^3) p_3^{6 k}
\right) \right \}\,.
\end{array}
}{detPsol2}

For each $k \in \natural_*$,
the solution depends on the real parameters $b_1,\ b_2$ and $b_3$, whose
values are restricted by the following condition:

\eqll{b_1\,b_2\ne 0\,.}{b}

In the rest of this section $k$ will be considered as fixed.

\begin{proposition}
For each $k$, the \wPms\ determined by the family S2 of proper solutions
form a one-parameter collection of distinct classes of equivalent \wPms.
\end{proposition}

\noindent{\em Proof:}
It will be advantageous to write the matrix $\wP(p)$ defined in
(\ref{matPsol2}) in a different (non-$F$-) basis.

With the following formal MIBT:

\eqll{\begin{array}{rcl}
p'_1 &=& {\displaystyle 1\over\displaystyle b_1^2} \left[
       p_1 +  b_1 (3 b_1 + 2 b_3) p_3^{3 k} - 6 b_1 p_2 p_3^k
      \right]  \\
&&\\
p'_2 &=&   {\displaystyle 1\over\displaystyle b_1}
      \left[ 2 p_2 - (3 b_1 + 2 b_3) p_3^{2 k} \right]  \\
&&\\
p'_3 &=& p_3\,,
\end{array}
}{MIBT1}
after setting

\eqll{(3 b_1 + 2 b_3)/b_1 = z}{z}
one obtains:

\eqll{\widehat F'(p') \,=\, {p'_1}^2 + {p'_2}^3\,,}{F'sol1}

\eqll{\begin{array}{rcl}
\wP'_{11}(p') &=&
{d_1}^{2} p'_2 {p'_3}^{k - 1}(-p'_1 + z p'_2 {p'_3}^{k})  \\
&&\\
\wP'_{12}(p') &=&
  - d_1 d_2 {p'_3}^{k - 1} (z p'_1 {p'_3}^{k} + {p'_2}^2)\\
&&\\
\wP'_{22}(p') &=&
   {d_2}^2 {p'_3}^{k - 1} (p'_1 - z p'_2 {p'_3}^{k}),
\end{array}
}{matP'sol2}
and
 %%%%%%%%%
\eqll{
{\rm det}(\wP'(p'))\, \,=\,
-4 {d_1}^{2} {d_2}^{2} {p'_3}^{k - 1} \left(z^2 {p'_3}^{3 k} +
p'_1 + (1 + z) p'_2 {p'_3}^{k}\right) ({p'_1}^2 + {p'_2}^3)\,.
}{detPxsol2}

We are left therefore with a unique free parameter $z$. A direct check shows
that \wPms\ corresponding to different values of $z$ cannot be related
by MIBT's; the parameter $z$ labels, therefore, the elements
of a one-parameter collection of non-equivalent \wPms, each
representing a class of equivalent \wPms.   \scat

Let us now prove that

\begin{proposition}
The proper solutions of the family S2 determine only one class of equivalent
A$\wP$M's. A representative \wPm\ is defined by (\ref{matP'sol2}) for $z=0$.
\end{proposition}

\noindent{\em Proof:}
It will be sufficient to analyse the representative matrices defined in
(\ref{matP'sol2}).

The algebraic variety $\cal Z$ determined by the relation $\widehat
F'(p')$ can be characterized by means of the following parametric equations:

\begin{eqnarray}
p'_1 &=& t^3 \nonumber \\
&& \\
p'_2 &=& -t^2\,.\nonumber
\end{eqnarray}

For each fixed value of $z$, let us determine the semi-positivity
domain ${\cal R}_{1}^{(\geq)}$ of $\wP'(p')$ in $\Pi \cap \cal Z$.

An immediate application of Theorem \ref{T2} shows that the region ${\cal
R}_1$ where $\wP'(t^3,-t^2,1)\,\ge\, 0$ and has rank 2 is determined by the
condition:

\eqll{t^2(z + t - t^2) \,>\, 0\,,}{r2}
while the region  where $\wP'(t^3,-t^2,1)\,\ge\, 0$  and has
rank 1 is determined by the condition

\eqll{t\,=\,0\,,\qquad {\rm or}\qquad  t\,=\, {1\over 2}\left(1\pm \sqrt{1+ 4
z}\right)\,. }{r1}
Therefore, condition {\em i)} of Lemma~\ref{Remark} is satisfied
if and only if $z=0$.
For $z\,=\,0$:
$${\cal R}_{1}^{(\geq)}\,=\,
\left \{ (p'_1,p'_2) \,\mid\,
p'_1=t^3,\, p'_2= -t^2,\; 0\,\le t\le 1 \right \}.$$
%%%%%%%%%   !!!!!!!!!!!!!!!!!    %%%%%%%%%%%%%

In order to check the boundary conditions at the singular strata (item {\em
ii)}
of Lemma~\ref{Remark}) in the case $z=0$, in Table \ref{table2}
we have reported the equations defining the singular primary
strata $\sigma^{(i)}$ of ${\cal R}^{(\geq)}$ and a possible choice for the
generators of the associated ideals ${\cal I}(\sigma^{(i)})$.
At this point it is easy to check that the boundary conditions
are satisfied at all the singular primary strata,
denoted by $\sigma^{(1)}$ and $\sigma^{(2)}$ in Fig.~2.
\scat

%%%%%%%%%%%%%%%%%%%%%%%%%%%
\begin{table}
\begin{center}
\begin{tabular}{||c|c|c|c|c||} \hline \hline
stratum    &  $t$  &defining equations & generators & boundary conditions
\\  \hline
 &  & $p'_1\,={p'_3}^{3k}$ & $f^{(1)}\,=\,{p'_1}- {p'_3}^{3k}$
 &
 \\
$\sigma^{(1)}$ &  $1$  &   &    &   satisfied \\
   &       &  $ p'_2=-{p'_3}^{2k}$ &    $f^{(2)}\,=\,p'_2 + {p'_3}^{2k}$
&
  \\     \hline
   &    & $p'_1\,=0 $ & $f^{(1)}\,=\,p'_1$ &
    \\
$\sigma^{(2)}$ &  $0$ &    &     & satisfied  \\
  &  & $p'_2\,=\,0$ & $f^{(2)}\,=\,p'_2 $ &
 \\
   \hline \hline
\end{tabular}

\end{center}
\caption{Defining equations for the primary strata, generators and
boundary conditions of solution S2 for $z=0$.}
\label{table2}
\end{table}
 %%%%%%%%%%%%%%%%%%%%%%%%%%%%%%%%%%%%%%%%%%%%%%%%%%%%

\begin{figure}
\centerline{ \epsfxsize=6.5cm \epsffile{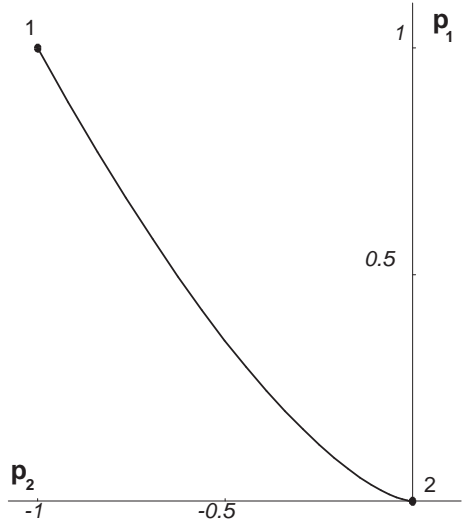} }
\caption{Stratification of ${\cal R}_{1}^{(\geq)}$ for solution S2
in the case $z=0$. The numbers $j$, $j=1,2\;$, label the singular
primary strata $\sigma^{(j)}$ defined in Table 3.}
\end{figure}

%%%%%%%%%%%%%%%    Figura  2 %%%%%%%%%%%%%%%%

%%%%%%%%%%%%%%%%%%%%%%%%%%%%%%%%%%%%%%%%%%%%%%%%%%%
 %
%

%%%%%%%%%%%%%%%%%%%%%%%%%%%%%%%%%%%%%%%%%%%%%%%%%%
\subsubsection{Solution S3}
The family S3 of proper solutions is found in correspondence with the degrees

\eqll{d_1 \,=\, d_2 \,=\, k\, \in\natural,\quad k\,\ge
\,2}{degSol3}
and for
\eqll{D\, =\, 0\,;\qquad n\, =\, 2\,.}{enne}
In the rest of this section, $k$ will be considered as fixed.

For the unknown polynomial functions one finds the following expressions:

$$L_1 \,=\, 0,\qquad L_2 \,=\, 0\,;$$

\eqll{\widehat F(p) \,\,= {\displaystyle\frac{1}{ 4 c_2}}
\left(-4 c_1 c_2 {p_3}^k + 4 c_2
p_1^2 + 4 c_2 c_3 p_1 p_2 + 4 c_1 p_2^2 + c_2 c_3^2 p_2^2\right);
}{Fsol3}

\eqll{\begin{array}{rcl}
\wP _{11}(p) &=&
{\displaystyle \frac{{d_1}^2}{4}} (4 c_1 + c_2 c_3^2)p_3^{k - 1} \\
&& \\
\wP_{12}(p) &=&
   - {\displaystyle \frac{d_1 d_2}{2}}c_2 c_3 p_3^{k - 1}   \\
&& \\
\wP _{22}(p) &=&
     {d_2}^2 c_2 p_3^{k - 1}\,,
\end{array}
}{matPsol3}
where $c_1,\ c_2$ and $c_3$ are real parameters, satisfying the condition:

\eqll{c_1\,c_2\ \ne\ 0\,.}{c1c2}

\begin{proposition}
For each fixed $k$, the \wPms\ determined by the family S3 of proper solutions
form four distinct classes of equivalent \wPms.
\end{proposition}

\noindent{\em Proof:}
The parameters $c_3$, $|c_{1}|$ and $|c_{2}|$
entering the definition of $\wP (p)$ in
(\ref{matPsol3}) can be eliminated by means of the following formal MIBT:

\eqll{\begin{array}{rcl}
p'_1 &=& |c_1|^{-1/2}\left(p_1 + c_3 p_2/2\right)  \\
&& \\
p'_2 &=& |c_2|^{-1/2}p_2   \\
&& \\
p'_3  &=& p_3\,.
\end{array}
}{MIBTsol2}
After setting
$$\epsilon_{i} = \mbox{sign}(c_i)\,, \qquad i=1,2\,,$$
in the new $F$-basis the non fixed elements of the \wPm\ defined in
(\ref{matPsol3}) assume the following simple form:

\eqll{\begin{array}{rcl}
\wP '_{11}(p') &=&
{d_1}^{2} \epsilon_{1} {p'_3}^{k - 1}  \\
&& \\
\wP' _{12}(p') &=&  0 \\
&& \\
\wP' _{22}(p') &=& {d_2}^{2} \epsilon_{2} {p'_3}^{k - 1}\,.
\end{array}
}{matP'sol3}
From (\ref{matP'sol3})
\eqll{{\rm det}\, \wP '(p') = 4 {d_1}^{2} {d_2}^2 \,
\epsilon_{1}\epsilon_{2}\left({p'_3}^{k} - \epsilon_{1}{p'_1}^2 - \epsilon_{2}
{p'_2}^2 \right)\,=\, -\epsilon_{2}\, \widehat F(p').
}{detmatP'}

It is now trivial to realize that the 4 matrices defined in (\ref{matP'sol3})
for $\epsilon_1=\pm 1$ and $\epsilon_2=\pm 1$ cannot be related by a MIBT.
For each fixed $k$, we are left therefore with only $4$ distinct classes of
equivalent \wPms.  \scat

\begin{proposition}
The proper solutions of the family S3 determine only one class of equivalent
A$\wP$M's. A representative \wPm\ is defined by (\ref{matP'sol3}) for
$\epsilon_1=\epsilon_2=1$.
\end{proposition}

\noindent{\em Proof:}
From (\ref{matP'sol3}) and (\ref{detmatP'}) it is easy to realize that the
semipositivity domain ${\cal R}_{1}^{(\geq)}$ of $\wP'(p')$ in $\Pi \cap {\cal
Z}$ is a $1$-dimensional connected semi-algebraic set only if
$\epsilon_1=\epsilon_2=1$.  In this case, ${\cal R}^{(\geq)}$ is the surface
of the upper half of a cone, and ${\cal R}_{1}^{(\geq)}$ is the unit circle
(see \cite{021}, Fig.~6, pag.  332).

It is trivial to check that the conditions listed in Lemma \ref{Remark}
are satisfied for $\epsilon_1=\epsilon_2=1$. Moreover, since the unique
singular stratum is the origin of $\real^q$, the boundary conditions are
certainly satisfied.
We conclude therefore that only for $\epsilon_1=\epsilon_2=1$ the
$P\widehat{P} M$'s defined in (\ref{matP'sol3}) are {\em allowable}. \scat

An interesting fact to note is that the $A \widehat{P} M$'s we have found have
the same form as the $A \wP M$'s of r-type $(3,3)$ of the solution of class
$I(1,1)$ reported in \cite{682}.

\subsection{Generating groups}
As stressed in \S 3 the \wPm\ generated by a CLG of r-type $(3,2)$ must
be equivalent to an A$\wP$M of r-type (3,2). The
results reported in the previous section show that it must be equivalent to a
matrix in one of the families S2 or S3.
As just pointed out, we cannot be sure, a priori, that the converse holds true
too, {\em i.e.,}\ that each of the matrices in the families S2 and S3 are
necessarily generated by a CLG. In the next two subsections we shall
study the problem in greater detail, separately for the solutions of the
families S2 and S3. The analysis will lead us to discover that a further
selection has to be done.

\subsubsection{Allowable solutions of the family S2}
 For each of the solutions of the family S2
the existence of a generating group can be excluded on the basis of
the following proposition:

\begin{proposition} \label{rimarco}
{\rm
If $f_1(p)$ and $f_2(p)$ are $w$-homogeneous prime polynomials and $\widehat{F}(p) =
f_1(p)^{n_1} - f_2(p)^{n_2}$, $n_1, n_2 >1,\;\; n_1, n_2 \in \natural$, then,
$\widehat{F}(p)$ cannot define a
basic relation among the elements of a minimal integrity basis of
a compact linear group of r-type $(q, q-1)$.
}
\end{proposition}

\noindent{\em Proof:}
Let us assume, in fact,  that $\{p_1(x), \dots , p_q(x)\}$ is a MIB of a compact linear
group of r-type $(q, q-1)$ and that the polynomial $\widehat{F}(p)$, defined in the
statement, defines the basic relation.  Then

\eqll{F(p(x))\,=\, f_1(p(x))^{n_1} - f_2(p(x))^{n_2}\, =\, 0}{rel}
is an identity in $x$ and should be the lowest degree relation
among the $p_a$'s. Let us assume, without loss of generality, that
$n_2\, \ge \, n_1 (\,>\, 1)$. Then, Eq.~(\ref{rel}) shows that
$\left[ f_1(p(x))/f_2(p(x)) \right]^{n_1}$, and consequently $f_1(p(x))/f_2(p(x))$, is to
be a $G$-invariant homogeneous polynomial $h(x)$:

\eqll{f_1(p(x))\,=\,h(x) f_2(p(x)).}{rel1}
Now, the degree of $h(x)$ is certainly lower than the degree of $F(p(x))$,
therefore there is a unique $w$-homogeneous polynomial $\hat h(p)$ such that
$\hat h(p(x)) = h(x)$, for all $x\in \real^n$ and

\eqll{\left.\left( f_1(p) - \hat h(p) f_2(p)\right)\right|_{\cal Z}\ =\
0\,.}{000}
Since the weight of the polynomial $f_1(p) - \hat h(p) f_2(p)$ is lower than
the weight of $F(p)$, equation (\ref{000}) can be extended to the whole of
$\real^q$. This contradicts
the assumption that $f_1(p)$ and $f_2(p)$ are prime polynomials.   \scat

\vskip3truemm
Proposition \ref{rimarco} suggests the following model for the generation of
proper solutions of class S2.

Let $q(x)=(q_1(x), q_2(x))$ define a MIB of a CLG, with degrees $(2k, 2)$. It is not
restrictive to assume that the associated \wPm\ has the following form
\cite{681}:

\eqll{\widehat{Q} (q_1, q_2)\,=\, \left(
\begin{array}{rr}
k^2 q_2^{k - 1}  &  2 k q_1 \\
2 k q_1        & 4 q_2
\end{array} \right). }{matQ}

Let us now define the following non-regular set $\{p\}$ of homogeneous
polynomial invariants:

\eqll{p_\alpha(x)\,=\,f_\alpha(q(x)),\quad \alpha = 1,2;\qquad
p_3(x)\,=\,q_2(x),}{basep}
where the $f_\alpha$'s are $w$-homogeneous polynomials.

The non fixed elements of the matrix $P(x)$, associated according to
(\ref{matP}) to the invariants $\{p_1(x), p_2(x),p_3(x)\}$, can be written in
the following form:

\eqll{ P_{\alpha\beta}(x)\,=\, \scalarp{\partial f_\alpha (q(x))}{\widehat
Q(q(x)) \partial f_\beta (q(x))},\qquad \alpha,\beta = 1,2.
}{matP1}
For general $f_1(q)$ and $f_2(q)$,
the set $\{ p(x) \}$ is not an integrity basis and the $P_{\alpha \beta}(x)$
cannot be expressed in terms of polynomials
in $p_1(x), p_2(x)$ and $p_3(x)$.
When this is possible, i.e., when a matrix $\wP(p(x))$ exists such that
$P(x)\,=\,\wP(p(x))$, then $\{ p(x) \}$ will be called
a {\em pseudo integrity basis}
(abbreviated in PIB).

We shall limit ourselves to determine those PIB's which are
relevant to the interpretation of the proper solutions of class S2.

A simple calculation shows that the following mono-parametric family of
couples of $w$-homogeneous polynomials $(f_1(q_1, q_2),f_2(q_1, q_2))$
of weights $d_1 = 6 k$, $d_2 = 4 k$ give rise to PIB's $\{p\}$:

\eqll{\begin{array}{rcl}
p_1&=& f_1(q_1, q_2) \,=\, q_1^3 - 3 s^2 q_1 q_2^{2 k} \\
&& \\
p_2&=&f_2(q_1, q_2) \,=\, q_1^2 + 2 s q_1 q_2^k\\
&& \\
p_3&=&q_2.
\end{array}
}{effee}

By eliminating $q_1$ and $q_2$ from the three equations in (\ref{effee}), the
following relation among the $p_a$'s is easily found:

\eqll{\widehat{F}(p)=\, \frac {1}{16}\left(
16 p_1^2 - 16 p_2^3 + 48 s p_1 p_2 p_3^k +
    24 s^2 p_2^2 p_3^{2 k} + 4 s^3 p_1 p_3^{3 k} +
    3 s^4 p_2 p_3^{4 k}\right).}{EFFE}

The \wPm\  associated to the PIB $\{p\}$ has the following non
fixed elements:

%%%%%%%%%%%%% %%%%%%%%%%%%%%%%%%%%%%%%%%%%%%%%%%%%%%%%%%%

\eqll{\begin{array}{rcl}
P_{11}(p) &=&
{\displaystyle \frac{d_1^2}{16}} p_3^{k -1}
      \left(-16 s^2 p_2^2 p_3^k +
    32 s^3 p_1 p_3^{2 k} +
    24 s^4 p_2 p_3^{3 k} +
    16 p_2^2 p_3^k +  \right.   \\
&&  \left.
    - 32 s p_1 p_3^{2 k} - 24 s^2 p_2 p_3^{3 k}  +
    s^4 p_3^{5 k}  \right)  \\
&&\\
P_{12}(p) &=&
{\displaystyle \frac{d_1 d_2}{8}} p_3^{k -1}
      \left(4 s p_2^2 - 12 s^2 p_1 p_3^k -
    9 s^3 p_2 p_3^{2 k}  +
    8 p_1 p_3^k  +
    4 s p_2 p_3^{2 k}  -
    s^3 p_3^{4 k}  \right) \\
&&\\
P_{22}(p) &=&
{\displaystyle \frac{d_2^2}{4}} p_3^{k -1}
    \left(4 s p_1 + 3 s^2 p_2 p_3^k +
    4 p_2 p_3^k  +
      s^2 p_3^{3 k} \right)\,.
\end{array}
}{matP2}
With the following MIBT:

\eqll{\begin{array}{rcl}
p'_1 &=& {\displaystyle\frac{1}{8 s^3}}\left(8 p_1 + 12 s p_2 p_3^k +
        s^3 p_3^{3 k}\right) \\
&&\\
p'_2 &=& -{\displaystyle\frac{1}{4 s^2}}\left(4 p_2 + s^2 p_3^{2 k}\right)
\\
&&\\
p'_3 &=& p_3,
\end{array}
}{MMIBT}
the matrix $\wP(p)$ is changed into the \wPm\  of
Eq.~(\ref{matP'sol2}), provided that

$$z\,=\, -{\displaystyle\frac{s^2-4 }{4 s^2}}.$$

It would be easy to check that also the $P \wP M$'s of the family S1 can be
generated in an analogous way.

We shall conclude this section with a comment. From a strictly rational point
of view, the restriction on the possible form of the basic relation among the
elements of a MIB of a group of r-type $(q,q-1)$ should have been included in
the definition of $A\wP M$'s. We have preferred to introduce it a posteriori to
stress the fact that the necessity of this further condition has not been
suggested to us by known results in invariant theory, but by our \wPm\
approach to the study of orbit spaces of CLG's.

 \subsubsection{Allowable \wPms\ of the family S3}
Contrary to what happens for the $A \wP M$'s of the family S2, it is not
difficult to find a generating group of the $A \wP M$'s of the family S3.
According to (\ref{enne}) it has to be searched for among the linear groups
acting in 2-dimensional spaces.

Let us consider for instance
the group $\integer_{n}$ generated by the
following transformation of the complex plane:
$$ z'= \exp\left(- \imath \frac{2\pi}{n}\right)\, z\,, \qquad z=x_1+\imath
\,x_2 \in \complex \,. $$
It is then evident that a MIB for $G$ is the following one:
$$p_1 \,=\, \Im (z^n) ,\qquad p_2 \,=\, \Re (z^n)\,, \qquad
p_3\,=\,z z^{*} \,=\, x_1^2 +x_2^2.$$
The MIB is not regular, since its elements satisfy the following identity:

$$p_1(x)^2 \,=\, p_3(x)^n - p_2(x)^2,$$
which exactly corresponds to the solution $\widehat{F}(p)$ we have found in our
approach.
The image of the orbit space $\overline{{\cal S}_{1}}$ is the unit circle.
The group $\integer_{n}$ provides a $2$-dimensional representation of
the point group $C_{n}$, $n \geq 2$, that
is the cyclic group of rotations about an axis of the $n$-th order.

%%%%%%%%%%%%%%%%%%%%%%%
As already noted, the $A \wP M$'s of r-type $(3,2)$ of the solution S3
are equivalent to the $A \wP M$'s of r-type $(3,3)$ of class $I(1,1)$
reported in \cite{682}. Generating groups of the first element of the
family are, for instance, the linear groups $\mbox{SO}(n, \real)$ acting in
$\real^{n} \oplus \real^{n}$ for $n \geq 3$.

\section{Concluding remarks}
To conclude, we would like to stress the following more or less unexpected
facts emerging from our analysis of non-coregular CLG's with only one
relation among the elements of their MIB's (class ${\cal T}(3,2)$):

\begin{itemize}
\item Coregular and non-coregular groups may share the same \wPm.

\item There is only one mono-parametric discrete family of allowable
non-equivalent \wPms\ $\wP^{(k)}(p)$, $\natural\ni k\ge 2$, whose elements may
be generated by groups $G\in{\cal T}(3,2)$.  The degrees of the $p_a$'s are
$d_1\,=\,d_2\,=\,k\, \ge 2$, $d_3\,=\,2$ and, with a convenient choice of the
$p_a$'s, the basic relation can be written in the form $\widehat{F}^{(k)}(p)\,
=\, p_1^2 + p_2^2 - p_3^k$.

\item Every allowable \wPm\ of the family is generated by at least a
group $G\in {\cal T}(3, 2)$.

\item If the action of the groups is restricted to the unit sphere
$S^{(n-1)}$ of $\real^n$ (which is not essentially restrictive for what
concerns the characterization of the orbit space), all the orbit spaces
$S^{(n-1)}/G$, $G\in {\cal T}(3, 2)$ turn out to be isomorphic.
\end{itemize}

%%%%%%%%%%%%%%%%%%%%%%%%

\end{document}